\hsize=31pc
\vsize=49pc
\lineskip=0pt
\parskip=0pt plus 1pt
\hfuzz=1pt
\vfuzz=2pt
\pretolerance=2500
\tolerance=5000
\vbadness=5000
\hbadness=5000
\widowpenalty=500
\clubpenalty=200
\brokenpenalty=500
\predisplaypenalty=200
\voffset=-1pc
\nopagenumbers
\catcode`@=11
\newif\ifams
\amsfalse
%
%
%
%
\newfam\bdifam
\newfam\bsyfam
\newfam\bssfam
\newfam\msafam
\newfam\msbfam
\newif\ifxxpt
\newif\ifxviipt
\newif\ifxivpt
\newif\ifxiipt
\newif\ifxipt
\newif\ifxpt
\newif\ifixpt
\newif\ifviiipt
\newif\ifviipt
\newif\ifvipt
\newif\ifvpt
%
%
\def\headsize#1#2{\def\headb@seline{#2}%
                \ifnum#1=20\def\HEAD{twenty}%
                           \def\smHEAD{twelve}%
                           \def\vsHEAD{nine}%
                           \ifxxpt\else\xdef\f@ntsize{\HEAD}%
                           \def\m@g{4}\def\s@ze{20.74}%
                           \loadheadfonts\xxpttrue\fi
                           \ifxiipt\else\xdef\f@ntsize{\smHEAD}%
                           \def\m@g{1}\def\s@ze{12}%
                           \loadxiiptfonts\xiipttrue\fi
                           \ifixpt\else\xdef\f@ntsize{\vsHEAD}%
                           \def\s@ze{9}%
                           \loadsmallfonts\ixpttrue\fi
                      \else
                \ifnum#1=17\def\HEAD{seventeen}%
                           \def\smHEAD{eleven}%
                           \def\vsHEAD{eight}%
                           \ifxviipt\else\xdef\f@ntsize{\HEAD}%
                           \def\m@g{3}\def\s@ze{17.28}%
                           \loadheadfonts\xviipttrue\fi
                           \ifxipt\else\xdef\f@ntsize{\smHEAD}%
                           \loadxiptfonts\xipttrue\fi
                           \ifviiipt\else\xdef\f@ntsize{\vsHEAD}%
                           \def\s@ze{8}%
                           \loadsmallfonts\viiipttrue\fi
                      \else\def\HEAD{fourteen}%
                           \def\smHEAD{ten}%
                           \def\vsHEAD{seven}%
                           \ifxivpt\else\xdef\f@ntsize{\HEAD}%
                           \def\m@g{2}\def\s@ze{14.4}%
                           \loadheadfonts\xivpttrue\fi
                           \ifxpt\else\xdef\f@ntsize{\smHEAD}%
                           \def\s@ze{10}%
                           \loadxptfonts\xpttrue\fi
                           \ifviipt\else\xdef\f@ntsize{\vsHEAD}%
                           \def\s@ze{7}%
                           \loadviiptfonts\viipttrue\fi
                \ifnum#1=14\else
                \message{Header size should be 20, 17 or 14 point
                              will now default to 14pt}\fi
                \fi\fi\headfonts}
%
%
\def\textsize#1#2{\def\textb@seline{#2}%
                 \ifnum#1=12\def\TEXT{twelve}%
                           \def\smTEXT{eight}%
                           \def\vsTEXT{six}%
                           \ifxiipt\else\xdef\f@ntsize{\TEXT}%
                           \def\m@g{1}\def\s@ze{12}%
                           \loadxiiptfonts\xiipttrue\fi
                           \ifviiipt\else\xdef\f@ntsize{\smTEXT}%
                           \def\s@ze{8}%
                           \loadsmallfonts\viiipttrue\fi
                           \ifvipt\else\xdef\f@ntsize{\vsTEXT}%
                           \def\s@ze{6}%
                           \loadviptfonts\vipttrue\fi
                      \else
                \ifnum#1=11\def\TEXT{eleven}%
                           \def\smTEXT{seven}%
                           \def\vsTEXT{five}%
                           \ifxipt\else\xdef\f@ntsize{\TEXT}%
                           \def\s@ze{11}%
                           \loadxiptfonts\xipttrue\fi
                           \ifviipt\else\xdef\f@ntsize{\smTEXT}%
                           \loadviiptfonts\viipttrue\fi
                           \ifvpt\else\xdef\f@ntsize{\vsTEXT}%
                           \def\s@ze{5}%
                           \loadvptfonts\vpttrue\fi
                      \else\def\TEXT{ten}%
                           \def\smTEXT{seven}%
                           \def\vsTEXT{five}%
                           \ifxpt\else\xdef\f@ntsize{\TEXT}%
                           \loadxptfonts\xpttrue\fi
                           \ifviipt\else\xdef\f@ntsize{\smTEXT}%
                           \def\s@ze{7}%
                           \loadviiptfonts\viipttrue\fi
                           \ifvpt\else\xdef\f@ntsize{\vsTEXT}%
                           \def\s@ze{5}%
                           \loadvptfonts\vpttrue\fi
                \ifnum#1=10\else
                \message{Text size should be 12, 11 or 10 point
                              will now default to 10pt}\fi
                \fi\fi\textfonts}
%
%
\def\smallsize#1#2{\def\smallb@seline{#2}%
                 \ifnum#1=10\def\SMALL{ten}%
                           \def\smSMALL{seven}%
                           \def\vsSMALL{five}%
                           \ifxpt\else\xdef\f@ntsize{\SMALL}%
                           \loadxptfonts\xpttrue\fi
                           \ifviipt\else\xdef\f@ntsize{\smSMALL}%
                           \def\s@ze{7}%
                           \loadviiptfonts\viipttrue\fi
                           \ifvpt\else\xdef\f@ntsize{\vsSMALL}%
                           \def\s@ze{5}%
                           \loadvptfonts\vpttrue\fi
                       \else
                 \ifnum#1=9\def\SMALL{nine}%
                           \def\smSMALL{six}%
                           \def\vsSMALL{five}%
                           \ifixpt\else\xdef\f@ntsize{\SMALL}%
                           \def\s@ze{9}%
                           \loadsmallfonts\ixpttrue\fi
                           \ifvipt\else\xdef\f@ntsize{\smSMALL}%
                           \def\s@ze{6}%
                           \loadviptfonts\vipttrue\fi
                           \ifvpt\else\xdef\f@ntsize{\vsSMALL}%
                           \def\s@ze{5}%
                           \loadvptfonts\vpttrue\fi
                       \else
                           \def\SMALL{eight}%
                           \def\smSMALL{six}%
                           \def\vsSMALL{five}%
                           \ifviiipt\else\xdef\f@ntsize{\SMALL}%
                           \def\s@ze{8}%
                           \loadsmallfonts\viiipttrue\fi
                           \ifvipt\else\xdef\f@ntsize{\smSMALL}%
                           \def\s@ze{6}%
                           \loadviptfonts\vipttrue\fi
                           \ifvpt\else\xdef\f@ntsize{\vsSMALL}%
                           \def\s@ze{5}%
                           \loadvptfonts\vpttrue\fi
                 \ifnum#1=8\else\message{Small size should be 10, 9 or
                            8 point will now default to 8pt}\fi
                \fi\fi\smallfonts}
\def\F@nt{\expandafter\font\csname}
\def\Sk@w{\expandafter\skewchar\csname}
\def\@nd{\endcsname}
\def\@step#1{ scaled \magstep#1}
\def\@half{ scaled \magstephalf}
\def\@t#1{ at #1pt}
%
%
\def\loadheadfonts{\bigf@nts
\F@nt \f@ntsize bdi\@nd=cmmib10 \@t{\s@ze}%
\Sk@w \f@ntsize bdi\@nd='177
\F@nt \f@ntsize bsy\@nd=cmbsy10 \@t{\s@ze}%
\Sk@w \f@ntsize bsy\@nd='60
\F@nt \f@ntsize bss\@nd=cmssbx10 \@t{\s@ze}}
%
%
\def\loadxiiptfonts{\bigf@nts
\F@nt \f@ntsize bdi\@nd=cmmib10 \@step{\m@g}%
\Sk@w \f@ntsize bdi\@nd='177
\F@nt \f@ntsize bsy\@nd=cmbsy10 \@step{\m@g}%
\Sk@w \f@ntsize bsy\@nd='60
\F@nt \f@ntsize bss\@nd=cmssbx10 \@step{\m@g}}
%
%
\def\loadxiptfonts{%
\font\elevenrm=cmr10 \@half
\font\eleveni=cmmi10 \@half
\skewchar\eleveni='177
\font\elevensy=cmsy10 \@half
\skewchar\elevensy='60
\font\elevenex=cmex10 \@half
\font\elevenit=cmti10 \@half
\font\elevensl=cmsl10 \@half
\font\elevenbf=cmbx10 \@half
\font\eleventt=cmtt10 \@half
\ifams\font\elevenmsa=msam10 \@half
\font\elevenmsb=msbm10 \@half\else\fi
\font\elevenbdi=cmmib10 \@half
\skewchar\elevenbdi='177
\font\elevenbsy=cmbsy10 \@half
\skewchar\elevenbsy='60
\font\elevenbss=cmssbx10 \@half}
%
%
\def\loadxptfonts{%
\font\tenbdi=cmmib10
\skewchar\tenbdi='177
\font\tenbsy=cmbsy10
\skewchar\tenbsy='60
\ifams\font\tenmsa=msam10
\font\tenmsb=msbm10\else\fi
\font\tenbss=cmssbx10}%
%
%
\def\loadsmallfonts{\smallf@nts
\ifams
\F@nt \f@ntsize ex\@nd=cmex\s@ze
\else
\F@nt \f@ntsize ex\@nd=cmex10\fi
\F@nt \f@ntsize it\@nd=cmti\s@ze
\F@nt \f@ntsize sl\@nd=cmsl\s@ze
\F@nt \f@ntsize tt\@nd=cmtt\s@ze}
%
%
\def\loadviiptfonts{%
\font\sevenit=cmti7
\font\sevensl=cmsl8 at 7pt
\ifams\font\sevenmsa=msam7
\font\sevenmsb=msbm7
\font\sevenex=cmex7
\font\sevenbsy=cmbsy7
\font\sevenbdi=cmmib7\else
\font\sevenex=cmex10
\font\sevenbsy=cmbsy10 at 7pt
\font\sevenbdi=cmmib10 at 7pt\fi
\skewchar\sevenbsy='60
\skewchar\sevenbdi='177
\font\sevenbss=cmssbx10 at 7pt}%
%
%
\def\loadviptfonts{\smallf@nts
\ifams\font\sixex=cmex7 at 6pt\else
\font\sixex=cmex10\fi
\font\sixit=cmti7 at 6pt}
%
%
\def\loadvptfonts{%
\font\fiveit=cmti7 at 5pt
\ifams\font\fiveex=cmex7 at 5pt
\font\fivebdi=cmmib5
\font\fivebsy=cmbsy5
\font\fivemsa=msam5
\font\fivemsb=msbm5\else
\font\fiveex=cmex10
\font\fivebdi=cmmib10 at 5pt
\font\fivebsy=cmbsy10 at 5pt\fi
\skewchar\fivebdi='177
\skewchar\fivebsy='60
\font\fivebss=cmssbx10 at 5pt}
\def\bigf@nts{%
\F@nt \f@ntsize rm\@nd=cmr10 \@step{\m@g}%
\F@nt \f@ntsize i\@nd=cmmi10 \@step{\m@g}%
\Sk@w \f@ntsize i\@nd='177
\F@nt \f@ntsize sy\@nd=cmsy10 \@step{\m@g}%
\Sk@w \f@ntsize sy\@nd='60
\F@nt \f@ntsize ex\@nd=cmex10 \@step{\m@g}%
\F@nt \f@ntsize it\@nd=cmti10 \@step{\m@g}%
\F@nt \f@ntsize sl\@nd=cmsl10 \@step{\m@g}%
\F@nt \f@ntsize bf\@nd=cmbx10 \@step{\m@g}%
\F@nt \f@ntsize tt\@nd=cmtt10 \@step{\m@g}%
\ifams
\F@nt \f@ntsize msa\@nd=msam10 \@step{\m@g}%
\F@nt \f@ntsize msb\@nd=msbm10 \@step{\m@g}\else\fi}
\def\smallf@nts{%
\F@nt \f@ntsize rm\@nd=cmr\s@ze
\F@nt \f@ntsize i\@nd=cmmi\s@ze
\Sk@w \f@ntsize i\@nd='177
\F@nt \f@ntsize sy\@nd=cmsy\s@ze
\Sk@w \f@ntsize sy\@nd='60
\F@nt \f@ntsize bf\@nd=cmbx\s@ze
\ifams
\F@nt \f@ntsize bdi\@nd=cmmib\s@ze
\F@nt \f@ntsize bsy\@nd=cmbsy\s@ze
\F@nt \f@ntsize msa\@nd=msam\s@ze
\F@nt \f@ntsize msb\@nd=msbm\s@ze
\else
\F@nt \f@ntsize bdi\@nd=cmmib10 \@t{\s@ze}%
\F@nt \f@ntsize bsy\@nd=cmbsy10 \@t{\s@ze}\fi
\Sk@w \f@ntsize bdi\@nd='177
\Sk@w \f@ntsize bsy\@nd='60
\F@nt \f@ntsize bss\@nd=cmssbx10 \@t{\s@ze}}%
%
%
\def\headfonts{%
\textfont0=\csname\HEAD rm\@nd
\scriptfont0=\csname\smHEAD rm\@nd
\scriptscriptfont0=\csname\vsHEAD rm\@nd
\def\rm{\fam0\csname\HEAD rm\@nd
\def\sc{\csname\smHEAD rm\@nd}}%
\textfont1=\csname\HEAD i\@nd
\scriptfont1=\csname\smHEAD i\@nd
\scriptscriptfont1=\csname\vsHEAD i\@nd
\textfont2=\csname\HEAD sy\@nd
\scriptfont2=\csname\smHEAD sy\@nd
\scriptscriptfont2=\csname\vsHEAD sy\@nd
\textfont3=\csname\HEAD ex\@nd
\scriptfont3=\csname\smHEAD ex\@nd
\scriptscriptfont3=\csname\smHEAD ex\@nd
\textfont\itfam=\csname\HEAD it\@nd
\scriptfont\itfam=\csname\smHEAD it\@nd
\scriptscriptfont\itfam=\csname\vsHEAD it\@nd
\def\it{\fam\itfam\csname\HEAD it\@nd
\def\sc{\csname\smHEAD it\@nd}}%
\textfont\slfam=\csname\HEAD sl\@nd
\def\sl{\fam\slfam\csname\HEAD sl\@nd
\def\sc{\csname\smHEAD sl\@nd}}%
\textfont\bffam=\csname\HEAD bf\@nd
\scriptfont\bffam=\csname\smHEAD bf\@nd
\scriptscriptfont\bffam=\csname\vsHEAD bf\@nd
\def\bf{\fam\bffam\csname\HEAD bf\@nd
\def\sc{\csname\smHEAD bf\@nd}}%
\textfont\ttfam=\csname\HEAD tt\@nd
\def\tt{\fam\ttfam\csname\HEAD tt\@nd}%
\textfont\bdifam=\csname\HEAD bdi\@nd
\scriptfont\bdifam=\csname\smHEAD bdi\@nd
\scriptscriptfont\bdifam=\csname\vsHEAD bdi\@nd
\def\bdi{\fam\bdifam\csname\HEAD bdi\@nd}%
\textfont\bsyfam=\csname\HEAD bsy\@nd
\scriptfont\bsyfam=\csname\smHEAD bsy\@nd
\def\bsy{\fam\bsyfam\csname\HEAD bsy\@nd}%
\textfont\bssfam=\csname\HEAD bss\@nd
\scriptfont\bssfam=\csname\smHEAD bss\@nd
\scriptscriptfont\bssfam=\csname\vsHEAD bss\@nd
\def\bss{\fam\bssfam\csname\HEAD bss\@nd}%
\ifams
\textfont\msafam=\csname\HEAD msa\@nd
\scriptfont\msafam=\csname\smHEAD msa\@nd
\scriptscriptfont\msafam=\csname\vsHEAD msa\@nd
\textfont\msbfam=\csname\HEAD msb\@nd
\scriptfont\msbfam=\csname\smHEAD msb\@nd
\scriptscriptfont\msbfam=\csname\vsHEAD msb\@nd
\else\fi
\normalbaselineskip=\headb@seline pt%
\setbox\strutbox=\hbox{\vrule height.7\normalbaselineskip
depth.3\baselineskip width0pt}%
\def\sc{\csname\smHEAD rm\@nd}\normalbaselines\bf}
%
%
\def\textfonts{%
\textfont0=\csname\TEXT rm\@nd
\scriptfont0=\csname\smTEXT rm\@nd
\scriptscriptfont0=\csname\vsTEXT rm\@nd
\def\rm{\fam0\csname\TEXT rm\@nd
\def\sc{\csname\smTEXT rm\@nd}}%
\textfont1=\csname\TEXT i\@nd
\scriptfont1=\csname\smTEXT i\@nd
\scriptscriptfont1=\csname\vsTEXT i\@nd
\textfont2=\csname\TEXT sy\@nd
\scriptfont2=\csname\smTEXT sy\@nd
\scriptscriptfont2=\csname\vsTEXT sy\@nd
\textfont3=\csname\TEXT ex\@nd
\scriptfont3=\csname\smTEXT ex\@nd
\scriptscriptfont3=\csname\smTEXT ex\@nd
\textfont\itfam=\csname\TEXT it\@nd
\scriptfont\itfam=\csname\smTEXT it\@nd
\scriptscriptfont\itfam=\csname\vsTEXT it\@nd
\def\it{\fam\itfam\csname\TEXT it\@nd
\def\sc{\csname\smTEXT it\@nd}}%
\textfont\slfam=\csname\TEXT sl\@nd
\def\sl{\fam\slfam\csname\TEXT sl\@nd
\def\sc{\csname\smTEXT sl\@nd}}%
\textfont\bffam=\csname\TEXT bf\@nd
\scriptfont\bffam=\csname\smTEXT bf\@nd
\scriptscriptfont\bffam=\csname\vsTEXT bf\@nd
\def\bf{\fam\bffam\csname\TEXT bf\@nd
\def\sc{\csname\smTEXT bf\@nd}}%
\textfont\ttfam=\csname\TEXT tt\@nd
\def\tt{\fam\ttfam\csname\TEXT tt\@nd}%
\textfont\bdifam=\csname\TEXT bdi\@nd
\scriptfont\bdifam=\csname\smTEXT bdi\@nd
\scriptscriptfont\bdifam=\csname\vsTEXT bdi\@nd
\def\bdi{\fam\bdifam\csname\TEXT bdi\@nd}%
\textfont\bsyfam=\csname\TEXT bsy\@nd
\scriptfont\bsyfam=\csname\smTEXT bsy\@nd
\def\bsy{\fam\bsyfam\csname\TEXT bsy\@nd}%
\textfont\bssfam=\csname\TEXT bss\@nd
\scriptfont\bssfam=\csname\smTEXT bss\@nd
\scriptscriptfont\bssfam=\csname\vsTEXT bss\@nd
\def\bss{\fam\bssfam\csname\TEXT bss\@nd}%
\ifams
\textfont\msafam=\csname\TEXT msa\@nd
\scriptfont\msafam=\csname\smTEXT msa\@nd
\scriptscriptfont\msafam=\csname\vsTEXT msa\@nd
\textfont\msbfam=\csname\TEXT msb\@nd
\scriptfont\msbfam=\csname\smTEXT msb\@nd
\scriptscriptfont\msbfam=\csname\vsTEXT msb\@nd
\else\fi
\normalbaselineskip=\textb@seline pt
\setbox\strutbox=\hbox{\vrule height.7\normalbaselineskip
depth.3\baselineskip width0pt}%
\everymath{}%
\def\sc{\csname\smTEXT rm\@nd}\normalbaselines\rm}
%
%
\def\smallfonts{%
\textfont0=\csname\SMALL rm\@nd
\scriptfont0=\csname\smSMALL rm\@nd
\scriptscriptfont0=\csname\vsSMALL rm\@nd
\def\rm{\fam0\csname\SMALL rm\@nd
\def\sc{\csname\smSMALL rm\@nd}}%
\textfont1=\csname\SMALL i\@nd
\scriptfont1=\csname\smSMALL i\@nd
\scriptscriptfont1=\csname\vsSMALL i\@nd
\textfont2=\csname\SMALL sy\@nd
\scriptfont2=\csname\smSMALL sy\@nd
\scriptscriptfont2=\csname\vsSMALL sy\@nd
\textfont3=\csname\SMALL ex\@nd
\scriptfont3=\csname\smSMALL ex\@nd
\scriptscriptfont3=\csname\smSMALL ex\@nd
\textfont\itfam=\csname\SMALL it\@nd
\scriptfont\itfam=\csname\smSMALL it\@nd
\scriptscriptfont\itfam=\csname\vsSMALL it\@nd
\def\it{\fam\itfam\csname\SMALL it\@nd
\def\sc{\csname\smSMALL it\@nd}}%
\textfont\slfam=\csname\SMALL sl\@nd
\def\sl{\fam\slfam\csname\SMALL sl\@nd
\def\sc{\csname\smSMALL sl\@nd}}%
\textfont\bffam=\csname\SMALL bf\@nd
\scriptfont\bffam=\csname\smSMALL bf\@nd
\scriptscriptfont\bffam=\csname\vsSMALL bf\@nd
\def\bf{\fam\bffam\csname\SMALL bf\@nd
\def\sc{\csname\smSMALL bf\@nd}}%
\textfont\ttfam=\csname\SMALL tt\@nd
\def\tt{\fam\ttfam\csname\SMALL tt\@nd}%
\textfont\bdifam=\csname\SMALL bdi\@nd
\scriptfont\bdifam=\csname\smSMALL bdi\@nd
\scriptscriptfont\bdifam=\csname\vsSMALL bdi\@nd
\def\bdi{\fam\bdifam\csname\SMALL bdi\@nd}%
\textfont\bsyfam=\csname\SMALL bsy\@nd
\scriptfont\bsyfam=\csname\smSMALL bsy\@nd
\def\bsy{\fam\bsyfam\csname\SMALL bsy\@nd}%
\textfont\bssfam=\csname\SMALL bss\@nd
\scriptfont\bssfam=\csname\smSMALL bss\@nd
\scriptscriptfont\bssfam=\csname\vsSMALL bss\@nd
\def\bss{\fam\bssfam\csname\SMALL bss\@nd}%
\ifams
\textfont\msafam=\csname\SMALL msa\@nd
\scriptfont\msafam=\csname\smSMALL msa\@nd
\scriptscriptfont\msafam=\csname\vsSMALL msa\@nd
\textfont\msbfam=\csname\SMALL msb\@nd
\scriptfont\msbfam=\csname\smSMALL msb\@nd
\scriptscriptfont\msbfam=\csname\vsSMALL msb\@nd
\else\fi
\normalbaselineskip=\smallb@seline pt%
\setbox\strutbox=\hbox{\vrule height.7\normalbaselineskip
depth.3\baselineskip width0pt}%
\everymath{}%
\def\sc{\csname\smSMALL rm\@nd}\normalbaselines\rm}%
\everydisplay{\indenteddisplay
   \gdef\labeltype{\eqlabel}}%
%
%
\def\hexnumber@#1{\ifcase#1 0\or 1\or 2\or 3\or 4\or 5\or 6\or 7\or 8\or
 9\or A\or B\or C\or D\or E\or F\fi}
\edef\bffam@{\hexnumber@\bffam}
\edef\bdifam@{\hexnumber@\bdifam}
\edef\bsyfam@{\hexnumber@\bsyfam}
\def\undefine#1{\let#1\undefined}
\def\newsymbol#1#2#3#4#5{\let\next@\relax
 \ifnum#2=\thr@@\let\next@\bdifam@\else
 \ifams
 \ifnum#2=\@ne\let\next@\msafam@\else
 \ifnum#2=\tw@\let\next@\msbfam@\fi\fi
 \fi\fi
 \mathchardef#1="#3\next@#4#5}
\def\mathhexbox@#1#2#3{\relax
 \ifmmode\mathpalette{}{\m@th\mathchar"#1#2#3}%
 \else\leavevmode\hbox{$\m@th\mathchar"#1#2#3$}\fi}

\def\bi#1{{\fam\bdifam\relax#1}}
%
%
\ifams\input amsmacro\fi
%
%
\newsymbol\bitGamma 3000
\newsymbol\bitDelta 3001
\newsymbol\bitTheta 3002
\newsymbol\bitLambda 3003
\newsymbol\bitXi 3004
\newsymbol\bitPi 3005
\newsymbol\bitSigma 3006
\newsymbol\bitUpsilon 3007
\newsymbol\bitPhi 3008
\newsymbol\bitPsi 3009
\newsymbol\bitOmega 300A
\newsymbol\balpha 300B
\newsymbol\bbeta 300C
\newsymbol\bgamma 300D
\newsymbol\bdelta 300E
\newsymbol\bepsilon 300F
\newsymbol\bzeta 3010
\newsymbol\bfeta 3011
\newsymbol\btheta 3012
\newsymbol\biota 3013
\newsymbol\bkappa 3014
\newsymbol\blambda 3015
\newsymbol\bmu 3016
\newsymbol\bnu 3017
\newsymbol\bxi 3018
\newsymbol\bpi 3019
\newsymbol\brho 301A
\newsymbol\bsigma 301B
\newsymbol\btau 301C
\newsymbol\bupsilon 301D
\newsymbol\bphi 301E
\newsymbol\bchi 301F
\newsymbol\bpsi 3020
\newsymbol\bomega 3021
\newsymbol\bvarepsilon 3022
\newsymbol\bvartheta 3023
\newsymbol\bvaromega 3024
\newsymbol\bvarrho 3025
\newsymbol\bvarzeta 3026
\newsymbol\bvarphi 3027
\newsymbol\bpartial 3040
\newsymbol\bell 3060
\newsymbol\bimath 307B
\newsymbol\bjmath 307C
\mathchardef\binfty "0\bsyfam@31
\mathchardef\bnabla "0\bsyfam@72
\mathchardef\bdot "2\bsyfam@01
\mathchardef\bGamma "0\bffam@00
\mathchardef\bDelta "0\bffam@01
\mathchardef\bTheta "0\bffam@02
\mathchardef\bLambda "0\bffam@03
\mathchardef\bXi "0\bffam@04
\mathchardef\bPi "0\bffam@05
\mathchardef\bSigma "0\bffam@06
\mathchardef\bUpsilon "0\bffam@07
\mathchardef\bPhi "0\bffam@08
\mathchardef\bPsi "0\bffam@09
\mathchardef\bOmega "0\bffam@0A
\mathchardef\itGamma "0100
\mathchardef\itDelta "0101
\mathchardef\itTheta "0102
\mathchardef\itLambda "0103
\mathchardef\itXi "0104
\mathchardef\itPi "0105
\mathchardef\itSigma "0106
\mathchardef\itUpsilon "0107
\mathchardef\itPhi "0108
\mathchardef\itPsi "0109
\mathchardef\itOmega "010A
\mathchardef\Gamma "0000
\mathchardef\Delta "0001
\mathchardef\Theta "0002
\mathchardef\Lambda "0003
\mathchardef\Xi "0004
\mathchardef\Pi "0005
\mathchardef\Sigma "0006
\mathchardef\Upsilon "0007
\mathchardef\Phi "0008
\mathchardef\Psi "0009
\mathchardef\Omega "000A
%
%
\newcount\firstpage  \firstpage=1  
\newcount\jnl                      
\newcount\secno                    
\newcount\subno                    
\newcount\subsubno                 
\newcount\appno                    
\newcount\tabno                    
\newcount\figno                    
\newcount\countno                  
\newcount\refno                    
\newcount\eqlett     \eqlett=97    
\newif\ifletter
\newif\ifwide
\newif\ifnotfull
\newif\ifaligned
\newif\ifnumbysec
\newif\ifappendix
\newif\ifnumapp
\newif\ifssf
\newif\ifppt
\newdimen\t@bwidth
\newdimen\c@pwidth
\newdimen\digitwidth                    
\newdimen\argwidth                      
\newdimen\secindent    \secindent=5pc   
\newdimen\textind    \textind=16pt      
\newdimen\tempval                       
\newskip\beforesecskip
\def\beforesecspace{\vskip\beforesecskip\relax}
\newskip\beforesubskip
\def\beforesubspace{\vskip\beforesubskip\relax}
\newskip\beforesubsubskip
\def\beforesubsubspace{\vskip\beforesubsubskip\relax}
\newskip\secskip
\def\secspace{\vskip\secskip\relax}
\newskip\subskip
\def\subspace{\vskip\subskip\relax}
\newskip\insertskip
\def\insertspace{\vskip\insertskip\relax}
\def\sp@ce{\ifx\next*\let\next=\@ssf
               \else\let\next=\@nossf\fi\next}
\def\@ssf#1{\nobreak\secspace\global\ssftrue\nobreak}
\def\@nossf{\nobreak\secspace\nobreak\noindent\ignorespaces}
\def\subsp@ce{\ifx\next*\let\next=\@sssf
               \else\let\next=\@nosssf\fi\next}
\def\@sssf#1{\nobreak\subspace\global\ssftrue\nobreak}
\def\@nosssf{\nobreak\subspace\nobreak\noindent\ignorespaces}
\beforesecskip=24pt plus12pt minus8pt
\beforesubskip=12pt plus6pt minus4pt
\beforesubsubskip=12pt plus6pt minus4pt
\secskip=12pt plus 2pt minus 2pt
\subskip=6pt plus3pt minus2pt
\insertskip=18pt plus6pt minus6pt%
\fontdimen16\tensy=2.7pt
\fontdimen17\tensy=2.7pt
%
%
\def\eqlabel{(\ifappendix\applett
               \ifnumbysec\ifnum\secno>0 \the\secno\fi.\fi
               \else\ifnumbysec\the\secno.\fi\fi\the\countno)}
\def\seclabel{\ifappendix\ifnumapp\else\applett\fi
    \ifnum\secno>0 \the\secno
    \ifnumbysec\ifnum\subno>0.\the\subno\fi\fi\fi
    \else\the\secno\fi\ifnum\subno>0.\the\subno
         \ifnum\subsubno>0.\the\subsubno\fi\fi}
\def\tablabel{\ifappendix\applett\fi\the\tabno}
\def\figlabel{\ifappendix\applett\fi\the\figno}
\def\gac{\global\advance\countno by 1}
%
%

\def\vfootnote#1{\insert\footins\bgroup
\interlinepenalty=\interfootnotelinepenalty
\splittopskip=\ht\strutbox 
\splitmaxdepth=\dp\strutbox \floatingpenalty=20000
\leftskip=0pt \rightskip=0pt \spaceskip=0pt \xspaceskip=0pt%
\noindent\smallfonts\rm #1\ \ignorespaces\footstrut\futurelet\next\fo@t}
%
%
\def\endinsert{\egroup
    \if@mid \dimen@=\ht0 \advance\dimen@ by\dp0
       \advance\dimen@ by12\p@ \advance\dimen@ by\pagetotal
       \ifdim\dimen@>\pagegoal \@midfalse\p@gefalse\fi\fi
    \if@mid \insertspace \box0 \par \ifdim\lastskip<\insertskip
    \removelastskip \penalty-200 \insertspace \fi
    \else\insert\topins{\penalty100
       \splittopskip=0pt \splitmaxdepth=\maxdimen
       \floatingpenalty=0
       \ifp@ge \dimen@=\dp0
       \vbox to\vsize{\unvbox0 \kern-\dimen@}%
       \else\box0\nobreak\insertspace\fi}\fi\endgroup}
%
%
%
\def\ind{\hbox to \secindent{\hfill}}
%
%

%
%

%
%
\def\indeqn#1{\alignedfalse\displ@y\halign{\hbox to \displaywidth
    {$\ind\@lign\displaystyle##\hfil$}\crcr #1\crcr}}
%
%
\def\indalign#1{\alignedtrue\displ@y \tabskip=0pt
  \halign to\displaywidth{\ind$\@lign\displaystyle{##}$\tabskip=0pt
    &$\@lign\displaystyle{{}##}$\hfill\tabskip=\centering
    &\llap{$\@lign\hbox{\rm##}$}\tabskip=0pt\crcr
    #1\crcr}}
\def\indenteddisplay#1$${\indispl@y{#1 }}
\def\indispl@y#1{\disptest#1\eqalignno\eqalignno\disptest}
\def\disptest#1\eqalignno#2\eqalignno#3\disptest{%
    \ifx#3\eqalignno
    \indalign#2%
    \else\indeqn{#1}\fi$$}
%
%

%
%

%
%

%
%

%
%

\def\ns{\noalign{\vskip-3pt}}

%

%
%
\def\bhbar{\rlap{\kern1pt\raise.4ex\hbox{\bf\char'40}}\bi{h}}
\def\case#1#2{{\textstyle{#1\over#2}}}

\def\d{{\rm d}}

\def\etal{{\it et al\/}\ }
\def\frac#1#2{{#1\over#2}}
\ifams
\def\lap{\lesssim}
\def\gap{\gtrsim}
\let\le=\leqslant

\let\leq=\leqslant

\let\geq=\geqslant
\else

\def\gap{\;\lower3pt\hbox{$\buildrel > \over \sim$}\;}%
\def\lap{\;\lower3pt\hbox{$\buildrel < \over \sim$}\;}\fi

\chardef\ii="10
\def\tqs{\hbox to 25pt{\hfil}}

\def\Bbbone{1\kern-.22em {\rm l}}
%
%
\def\rp{\raise8pt\hbox{$\scriptstyle\prime$}}
%
%
%
%

%
%
\def\[#1\]{\setbox0=\hbox{$\dsty#1$}\argwidth=\wd0
    \setbox0=\hbox{$\left[\box0\right]$}\advance\argwidth by -\wd0
    \left[\kern.3\argwidth\box0\kern.3\argwidth\right]}
%
%
\def\lsb#1\rsb{\setbox0=\hbox{$#1$}\argwidth=\wd0
    \setbox0=\hbox{$\left[\box0\right]$}\advance\argwidth by -\wd0
    \left[\kern.3\argwidth\box0\kern.3\argwidth\right]}
%

%
%

%
\def\pt(#1){({\it #1\/})}
\let\dsty=\displaystyle

%
%
\def\reactions#1{\vskip 12pt plus2pt minus2pt%
\vbox{\hbox{\kern\secindent\vrule\kern12pt%
\vbox{\kern0.5pt\vbox{\hsize=24pc\parindent=0pt\smallfonts\rm NUCLEAR
REACTIONS\strut\quad #1\strut}\kern0.5pt}\kern12pt\vrule}}}
%
%
\def\slashchar#1{\setbox0=\hbox{$#1$}\dimen0=\wd0%
\setbox1=\hbox{/}\dimen1=\wd1%
\ifdim\dimen0>\dimen1%
\rlap{\hbox to \dimen0{\hfil/\hfil}}#1\else
\rlap{\hbox to \dimen1{\hfil$#1$\hfil}}/\fi}
%
%
\def\textindent#1{\noindent\hbox to \parindent{#1\hss}\ignorespaces}
%
%
\def\opencirc{\raise1pt\hbox{$\scriptstyle{\bigcirc}$}}

\ifams
\def\opensqr{\hbox{$\square$}}

\def\opentridown{\hbox{$\triangledown$}}

\else
\def\opensqr{\vbox{\hrule height.4pt\hbox{\vrule width.4pt height3.5pt
    \kern3.5pt\vrule width.4pt}\hrule height.4pt}}

\def\opentridown{\raise1pt\hbox{$\scriptstyle\bigtriangledown$}}

\fi

%
%
\def\m@th{\mathsurround=0pt}
%
%
\def\cases#1{%
\left\{\,\vcenter{\normalbaselines\openup1\jot\m@th%
     \ialign{$\displaystyle##\hfil$&\rm\tqs##\hfil\crcr#1\crcr}}\right.}%
%
%
\def\oldcases#1{\left\{\,\vcenter{\normalbaselines\m@th
    \ialign{$##\hfil$&\rm\quad##\hfil\crcr#1\crcr}}\right.}
%
%
\def\numcases#1{\left\{\,\vcenter{\baselineskip=15pt\m@th%
     \ialign{$\displaystyle##\hfil$&\rm\tqs##\hfil
     \crcr#1\crcr}}\right.\hfill
     \vcenter{\baselineskip=15pt\m@th%
     \ialign{\rlap{$\phantom{\displaystyle##\hfil}$}\tabskip=0pt&\en
     \rlap{\phantom{##\hfil}}\crcr#1\crcr}}}
\def\ptnumcases#1{\left\{\,\vcenter{\baselineskip=15pt\m@th%
     \ialign{$\displaystyle##\hfil$&\rm\tqs##\hfil
     \crcr#1\crcr}}\right.\hfill
     \vcenter{\baselineskip=15pt\m@th%
     \ialign{\rlap{$\phantom{\displaystyle##\hfil}$}\tabskip=0pt&\enpt
     \rlap{\phantom{##\hfil}}\crcr#1\crcr}}\global\eqlett=97
     \global\advance\countno by 1}
%
%
\def\eq(#1){\ifaligned\@mp(#1)\else\hfill\llap{{\rm (#1)}}\fi}
\def\ceq(#1){\ns\ns\ifaligned\@mp\fi\eq(#1)\cr\ns\ns}
\def\eqpt(#1#2){\ifaligned\@mp(#1{\it #2\/})
                    \else\hfill\llap{{\rm (#1{\it #2\/})}}\fi}

%
%
\countno=1

\def\aleq{&\rm(\ifappendix\applett
               \ifnumbysec\ifnum\secno>0 \the\secno\fi.\fi
               \else\ifnumbysec\the\secno.\fi\fi\the\countno}
\def\noaleq{\hfill\llap\bgroup\rm(\ifappendix\applett
               \ifnumbysec\ifnum\secno>0 \the\secno\fi.\fi
               \else\ifnumbysec\the\secno.\fi\fi\the\countno}
\def\@mp{&}
\def\en{\ifaligned\aleq)\else\noaleq)\egroup\fi\gac}
\def\cen{\ns\ns\ifaligned\@mp\fi\en\cr\ns\ns}
\def\enpt{\ifaligned\aleq{\it\char\the\eqlett})\else
    \noaleq{\it\char\the\eqlett})\egroup\fi
    \global\advance\eqlett by 1}
\def\endpt{\ifaligned\aleq{\it\char\the\eqlett})\else
    \noaleq{\it\char\the\eqlett})\egroup\fi
    \global\eqlett=97\gac}
%
%
\def\CQG{{\it Class.\ Quantum Grav.}}




%
%

\def\PR{{\it Phys. Rev.}}

\headline={\ifodd\pageno{\ifnum\pageno=\firstpage\hfill
   \else\rrhead\fi}\else\lrhead\fi}
\def\rrhead{\textfonts\hskip\secindent\it
    \shorttitle\hfill\rm\folio}
\def\lrhead{\textfonts\hbox to\secindent{\rm\folio\hss}%
    \it\aunames\hss}
\footline={\ifnum\pageno=\firstpage \hfill\textfonts\rm\folio\fi}
\def\@rticle#1#2{\vglue.5pc
    {\parindent=\secindent \bf #1\par}
     \vskip2.5pc
    {\exhyphenpenalty=10000\hyphenpenalty=10000
     \baselineskip=18pt\raggedright\noindent
     \headfonts\bf#2\par}\futurelet\next\sh@rttitle}%
\def\title#1{\gdef\shorttitle{#1}
    \vglue4pc{\exhyphenpenalty=10000\hyphenpenalty=10000
    \baselineskip=18pt
    \raggedright\parindent=0pt
    \headfonts\bf#1\par}\futurelet\next\sh@rttitle}

\def\article#1#2{\gdef\shorttitle{#2}\@rticle{#1}{#2}}
\def\review#1{\gdef\shorttitle{#1}%
    \@rticle{REVIEW \ifpbm\else ARTICLE\fi}{#1}}
\def\topical#1{\gdef\shorttitle{#1}%
    \@rticle{TOPICAL REVIEW}{#1}}
\def\comment#1{\gdef\shorttitle{#1}%
    \@rticle{COMMENT}{#1}}
\def\note#1{\gdef\shorttitle{#1}%
    \@rticle{NOTE}{#1}}
\def\prelim#1{\gdef\shorttitle{#1}%
    \@rticle{PRELIMINARY COMMUNICATION}{#1}}
\def\letter#1{\gdef\shorttitle{Letter to the Editor}%
     \gdef\aunames{Letter to the Editor}
     \global\lettertrue\ifnum\jnl=7\global\letterfalse\fi
     \@rticle{LETTER TO THE EDITOR}{#1}}
\def\sh@rttitle{\ifx\next[\let\next=\sh@rt
                \else\let\next=\f@ll\fi\next}
\def\sh@rt[#1]{\gdef\shorttitle{#1}}
\def\f@ll{}
\def\author#1{\ifletter\else\gdef\aunames{#1}\fi\vskip1.5pc
    {\parindent=\secindent
     \hang\textfonts
     \ifppt\bf\else\rm\fi#1\par}
     \ifppt\bigskip\else\smallskip\fi
     \futurelet\next\@unames}
\def\@unames{\ifx\next[\let\next=\short@uthor
                 \else\let\next=\@uthor\fi\next}
\def\short@uthor[#1]{\gdef\aunames{#1}}
\def\@uthor{}
\def\address#1{{\parindent=\secindent
    \exhyphenpenalty=10000\hyphenpenalty=10000
\ifppt\textfonts\else\smallfonts\fi\hang\raggedright\rm#1\par}%
    \ifppt\bigskip\fi}
\def\jl#1{\global\jnl=#1}
\jl{0}%
\def\journal{\ifnum\jnl=1 J. Phys.\ A: Math.\ Gen.\
        \else\ifnum\jnl=2 J. Phys.\ B: At.\ Mol.\ Opt.\ Phys.\
        \else\ifnum\jnl=3 J. Phys.:\ Condens. Matter\
        \else\ifnum\jnl=4 J. Phys.\ G: Nucl.\ Part.\ Phys.\
        \else\ifnum\jnl=5 Inverse Problems\
        \else\ifnum\jnl=6 Class. Quantum Grav.\
        \else\ifnum\jnl=7 Network\
        \else\ifnum\jnl=8 Nonlinearity\
        \else\ifnum\jnl=9 Quantum Opt.\
        \else\ifnum\jnl=10 Waves in Random Media\
        \else\ifnum\jnl=11 Pure Appl. Opt.\
        \else\ifnum\jnl=12 Phys. Med. Biol.\
        \else\ifnum\jnl=13 Modelling Simulation Mater.\ Sci.\ Eng.\
        \else\ifnum\jnl=14 Plasma Phys. Control. Fusion\
        \else\ifnum\jnl=15 Physiol. Meas.\
        \else\ifnum\jnl=16 Sov.\ Lightwave Commun.\
        \else\ifnum\jnl=17 J. Phys.\ D: Appl.\ Phys.\
        \else\ifnum\jnl=18 Supercond.\ Sci.\ Technol.\
        \else\ifnum\jnl=19 Semicond.\ Sci.\ Technol.\
        \else\ifnum\jnl=20 Nanotechnology\
        \else\ifnum\jnl=21 Meas.\ Sci.\ Technol.\
        \else\ifnum\jnl=22 Plasma Sources Sci.\ Technol.\
        \else\ifnum\jnl=23 Smart Mater.\ Struct.\
        \else\ifnum\jnl=24 J.\ Micromech.\ Microeng.\
   \else Institute of Physics Publishing\
   \fi\fi\fi\fi\fi\fi\fi\fi\fi\fi\fi\fi\fi\fi\fi
   \fi\fi\fi\fi\fi\fi\fi\fi\fi}
\let\abs=\beginabstract

\let\endabs=\endabstract
\def\submitted{\ifppt\noindent\textfonts\rm Submitted to \journal\par
     \bigskip\fi}
\def\today{\number\day\ \ifcase\month\or
     January\or February\or March\or April\or May\or June\or
     July\or August\or September\or October\or November\or
     December\fi\space \number\year}
\def\date{\ifppt\noindent\textfonts\rm
     Date: \today\par\goodbreak\bigskip\fi}
%
%
\def\pacs#1{\ifppt\noindent\textfonts\rm
     PACS number(s): #1\par\bigskip\fi}
%

%
%
\def\section#1{\ifppt\ifnum\secno=0\eject\fi\fi
    \subno=0\subsubno=0\global\advance\secno by 1
    \gdef\labeltype{\seclabel}\ifnumbysec\countno=1\fi
    \goodbreak\beforesecspace\nobreak
    \noindent{\bf \the\secno. #1}\par\futurelet\next\sp@ce}
\def\subsection#1{\subsubno=0\global\advance\subno by 1
     \gdef\labeltype{\seclabel}%
     \ifssf\else\goodbreak\beforesubspace\fi
     \global\ssffalse\nobreak
     \noindent{\it \the\secno.\the\subno. #1\par}%
     \futurelet\next\subsp@ce}
\def\subsubsection#1{\global\advance\subsubno by 1
     \gdef\labeltype{\seclabel}%
     \ifssf\else\goodbreak\beforesubsubspace\fi
     \global\ssffalse\nobreak
     \noindent{\it \the\secno.\the\subno.\the\subsubno. #1}\null.
     \ignorespaces}
%

%
%
\def\numappendix#1{\ifappendix\ifnumbysec\countno=1\fi\else
    \countno=1\figno=0\tabno=0\fi
    \subno=0\global\advance\appno by 1
    \secno=\appno\gdef\applett{A}\gdef\labeltype{\seclabel}%
    \global\appendixtrue\global\numapptrue
    \goodbreak\beforesecspace\nobreak
    \noindent{\bf Appendix \the\appno. #1\par}%
    \futurelet\next\sp@ce}
\def\numsubappendix#1{\global\advance\subno by 1\subsubno=0
    \gdef\labeltype{\seclabel}%
    \ifssf\else\goodbreak\beforesubspace\fi
    \global\ssffalse\nobreak
    \noindent{\it A\the\appno.\the\subno. #1\par}%
    \futurelet\next\subsp@ce}
\def\@ppendix#1#2#3{\countno=1\subno=0\subsubno=0\secno=0\figno=0\tabno=0
    \gdef\applett{#1}\gdef\labeltype{\seclabel}\global\appendixtrue
    \goodbreak\beforesecspace\nobreak
    \noindent{\bf Appendix#2#3\par}\futurelet\next\sp@ce}
\def\Appendix#1{\@ppendix{A}{. }{#1}}
\def\appendix#1#2{\@ppendix{#1}{ #1. }{#2}}
\def\App#1{\@ppendix{A}{ }{#1}}
\def\app{\@ppendix{A}{}{}}
\def\subappendix#1#2{\global\advance\subno by 1\subsubno=0
    \gdef\labeltype{\seclabel}%
    \ifssf\else\goodbreak\beforesubspace\fi
    \global\ssffalse\nobreak
    \noindent{\it #1\the\subno. #2\par}%
    \nobreak\subspace\noindent\ignorespaces}
%
%
\def\@ck#1{\ifletter\bigskip\noindent\ignorespaces\else
    \goodbreak\beforesecspace\nobreak
    \noindent{\bf Acknowledgment#1\par}%
    \nobreak\secspace\noindent\ignorespaces\fi}
\def\ack{\@ck{s}}
\def\ackn{\@ck{}}
\def\n@ip#1{\goodbreak\beforesecspace\nobreak
    \noindent\smallfonts{\it #1}. \rm\ignorespaces}
\def\naip{\n@ip{Note added in proof}}
\def\na{\n@ip{Note added}}

%
%

%

%
%

%

%

\def\tablecont{\topinsert\global\advance\tabno by -1
    \tablecaption{(continued)}}
\def\tablecaption#1{\gdef\labeltype{\tablabel}\global\widefalse
    \leftskip=\secindent\parindent=0pt
    \global\advance\tabno by 1
    \smallfonts{\bf Table \ifappendix\applett\fi\the\tabno.} \rm #1\par
    \smallskip\futurelet\next\t@b}
\def\t@b{\ifx\next*\let\next=\widet@b
             \else\ifx\next[\let\next=\fullwidet@b
                      \else\let\next=\narrowt@b\fi\fi
             \next}
\def\widet@b#1{\global\widetrue\global\notfulltrue
    \t@bwidth=\hsize\advance\t@bwidth by -\secindent}
\def\fullwidet@b[#1]{\global\widetrue\global\notfullfalse
    \leftskip=0pt\t@bwidth=\hsize}
\def\narrowt@b{\global\notfulltrue}
\def\align{\catcode`?=13\ifnotfull\moveright\secindent\fi
    \vbox\bgroup\halign\ifwide to \t@bwidth\fi
    \bgroup\strut\tabskip=1.2pc plus1pc minus.5pc}
\def\endalign{\egroup\egroup\catcode`?=12}

%
%

%
%

%

%
%

%

\catcode`?=13
\def\lineup{\setbox0=\hbox{\smallfonts\rm 0}%
    \digitwidth=\wd0%
    \def?{\kern\digitwidth}%
    \def\\{\hbox{$\phantom{-}$}}%
    \def\-{\llap{$-$}}}
\catcode`?=12
%
%
\def\sidetable#1#2{\hbox{\ifppt\hsize=18pc\t@bwidth=18pc
                          \else\hsize=15pc\t@bwidth=15pc\fi
    \parindent=0pt\vtop{\null #1\par}%
    \ifppt\hskip1.2pc\else\hskip1pc\fi
    \vtop{\null #2\par}}}
\def\lstable#1#2{\everypar{}\tempval=\hsize\hsize=\vsize
    \vsize=\tempval\hoffset=-3pc
    \global\tabno=#1\gdef\labeltype{\tablabel}%
    \noindent\smallfonts{\bf Table \ifappendix\applett\fi
    \the\tabno.} \rm #2\par
    \smallskip\futurelet\next\t@b}
\def\inctabno{\global\advance\tabno by 1}
%
%

%

%
\def\figure#1{\figc@ption{#1}\bigskip}
\def\figc@ption#1{\global\advance\figno by 1\gdef\labeltype{\figlabel}%
   {\parindent=\secindent\smallfonts\hang
    {\bf Figure \ifappendix\applett\fi\the\figno.} \rm #1\par}}
%
%
\def\refHEAD{\goodbreak\beforesecspace
     \noindent\textfonts{\bf References}\par
     \let\ref=\rf
     \nobreak\smallfonts\rm}
\def\references{\refHEAD\parindent=0pt
     \everypar{\hangindent=18pt\hangafter=1
     \frenchspacing\rm}%
     \secspace}
\def\rf#1{\par\noindent\hbox to 21pt{\hss #1\quad}\ignorespaces}
\def\refjl#1#2#3#4{\noindent #1 {\it #2 \bf #3} #4\par}
\def\refbk#1#2#3{\noindent #1 {\it #2} #3\par}
%
%

%
%

%
%

%
%

%
\catcode`\@=12
%
%
\def\pptstyle{\ppttrue\headsize{17}{24}%
\textsize{12}{16}%
\smallsize{10}{12}%
\hsize=37.2pc\vsize=56pc
\textind=20pt\secindent=6pc}
%
%

%
%

%
%

%
\parindent=\textind

\input xref
\input epsf
%
%

%
%
\def\allinea#1{%
\null\,\vcenter{\openup1\jot \mathsurround=0pt
\tabskip=0pt\vskip-\smallskipamount\everycr={\noalign{\smallskip}}
\halign{%
&\hfil$\displaystyle{{}##{}}$&$\displaystyle{{}##{}}$\hfil\crcr
#1\crcr
\noalign{\vskip-\smallskipamount}}}\,}
%
\long\def\iniziaitem#1\fineitem
 {{\smallskip\parindent=1truecm
 \parskip=10truept #1\medskip}}
%
\def\nuovariga{\hfil\break}
\def\ListaFigure{\vfill\eject\global\appendixfalse\textfonts\rm
\everypar{}\null\bigskip}
\def\a{\alpha}
\def\be{\beta}
\def\de{\delta}
\def\g{\gamma}
\def\eps{\varepsilon}
\def\r{\varrho}
\def\ST{space--time}
\def\CdR{Regge Calculus}
\def\RG{General Relativity}
\def\phz{{\vphantom0}}
\def\phm{{\phantom-}}
\def\vphhalf{{\vphantom{\case12}}}
\def\bu{$\bullet$}
\def\sigmal{\sigma^{\vphantom{j}}_{{l'}^2}}
\def\sigmad{\sigma^{\vphantom{j}}_{d^2}}
\def\vom{v^{\vphantom{j}}_\Omega}

\def\hsmash#1{\setbox0=\hbox{$\scriptstyle#1$} \wd0=0pt \box0}
\def\valass#1{|#1\mkern 1.5mu |}
\def\ugual{\,\mathop=\limits^{\rm def}\,}
\def\tonde#1{$\left(\hbox{#1}\vphhalf\right)$}
\def\tondemat#1{\!\left(#1\vphhalf\right)}
\def\apritonda{$\left(\vphhalf\right.\!$}
\def\chiuditonda{$\!\left.\vphhalf\right)$}
\def\plico{\vrule height 12.7pt width 0pt}
\def\vu{\skew{-2}\vec u}
\def\10#1#2{#1\cdot10^{#2}}
\def\funzR{\left(\smash{1-\dot R^2}\vphantom{R^2}\right)}
\def\sqlk#1#2{{}^{(k)}#1 s_{#2}}
%
\def\Brewin{Brewin (1987)}
\def\BrewinG{Brewin and Gentle}
\def\Williams{Collins and Williams (1973)}
\def\Barrett{Barrett \etal (1997)}
\def\Barret{Barrett (1986)}
\def\MillerG{Gentle and Miller (1998)}
\def\Coxeter{Coxeter (1973)}
\def\Tuckey{Tuckey (1993)}
\def\Sorkin{Sorkin (1975)}
\pptstyle          
\title{The Friedmann universe of dust by Regge Calculus: study of its ending
point}%
[The Friedmann universe by Regge Calculus]
\jl{6}

\author{A De Felice%
\footnote{\dag}{e-mail: \tt antonio@astr13pi.difi.unipi.it}
and E Fabri\ddag
\footnote{\S}{e-mail: \tt fabri@df.unipi.it}%
}[A De Felice and E Fabri]

\address{\ddag\ Department of Physics,
University of Pisa, p.zza Torricelli 2, 56126 Pisa PI, Italy}

\abs
We develop an evolution scheme, based on Sorkin algorithm, to evolve the most
complex regular tridimensional polytope, the 600-cell. This application has
been already studied before and all authors found a stop point for the
evolution of the spatial section. In our opinion a clear and satisfactory
meaning to this behaviour has not been given. In this paper we propose a
reason why the evolution of the 600-cell stops when its volume is still
far from 0. We find that the 600-cell meets a causality-breaking
singularity of \ST. We study the nature of this singularity by embedding the
600-cell into a five-dimensional Lorentzian manifold. We fit 600-cell's
evolution with a continuos metric and study it as a solution of Einstein
equations.
\endabs

\pacs{04.20.-q, 04.25.Dm}

\submitted

\date

\section{Introduction}
A key test of \CdR\ which has been studied many times in literature is the
Friedmann universe of dust.

The first papers on the Friedmann universe did not use simplicial complexes.
The authors made use of non-simplicial blocks for which they had to give some
other information besides the edge lengths. This information was
deduced by symmetry.
\Williams\ used three tridimensional regular polytopes to describe the spatial
sections with topology $S^3$ of the Friedmann universe of dust.
Then \Brewin\ wrote in the discrete formalism of \CdR\ the expression of the
action. Following \Barrett, for a regular 3-polytope having $n$
vertices, matter is placed on the timelike edges joining a vertex $i$ of the
polytope with its evolute $i'$ of the next spatial section, finding:
$$
S=\frac 1{8\pi }\sum_b\eps _b\,A_b-\frac Mn\sum_il_{ii'},\label{azione}$$
where the first sum is over the bones $b$ of the manifold and $\eps_b$ is the
defect of the bone $b$ with area $A_b$. The second sum is done over every
vertex $i$ of the first spatial section. Since
a regular polytope having $n=120$ has been chosen as spatial section, a mass
$M/120$ is placed in each
vertex $i$. Besides the regular polytopes, Brewin also used other non regular
tridimensional polytopes. The evolution of all his polytopes, using Regge
equations, stopped before the volume of spatial 3-section became zero: a
point was reached where the equations had no solution. He saw that this point
approaches the limit value (spatial section of null radius) as the polytope
approaches the 3-sphere of the continuum solution. To overcome the obstacle,
Brewin continued the evolution reverting to differential equations.

The first authors who used a simplicial complex for this metric were \Barrett.
They evolved the Friedmann universe of dust using the Regge equations found
by extremizing the action \ref{azione}:
$$\sum_k\eps _{ijk}\,\frac{\partial A_{ijk}}{%
\partial l_{ij}}=\frac \pi {15}\,M\,\delta_{i'j},$$
where the sum is done over all vertices $k$ joined to the edge $[ij]$ of
length $l_{ij}$. The bone $[ijk]$ has $\eps_{ijk}$ as defect and $A_{ijk}$
as area. They made use of the evolutive scheme found by \Sorkin, but in our
opinion their simplicial complex is not built correctly (see subsubsection
\ref{sezvertici}). However they too
found the point of stop and saw that its position is independent of
the timelike interval between two consecutive spatial sections. The authors
advanced the idea that the evolution should be continued by making spacelike
the edges of the lattice which were timelike before the stop. However they did
not pursue the idea.

In the present paper we propose to solve the problem of the stop of the
evolution and to understand the meaning of the ending point.
The paper is organized as follows. In section \ref{secEmb} we embed the
Robertson--Walker metric in a five-dimensional Lorentzian space. We will find
it useful in order to understand the problems related to the evolution of our lattice.
Then we describe our lattice and carry out the first evolution in order to
create the initial simplicial sandwich (section~\ref{600topo}).
Section~\ref{evoluzioni} contains the evolution of the initial sandwich and
the results obtained. The description of a correct evolution algorithm for our
simplicial complex based on Regge equations and the Sorkin evolutive scheme is
also presented here. The nature of the singularities belonging to a general
closed Robertson--Walker metric is discussed in section~\ref{analisi}. In
section~\ref{problema} we propose a deeper reason why our simplicial
approximation of the Friedmann universe ceases to exist before the volume of
the spatial 3-section vanishes. We find that the stop condition is caused by a
novel singularity of the metric. We fit the simplicial solution with a
continuous metric and study its stress-energy tensor (section~\ref{nuovo}).
Finally, our conclusions are reported in section~\ref{conclusioni}.

\section{Embedding into a 5-dimensional space\label{secEmb}}
We shall find convenient to embed a \ST\ with closed Robertson--Walker
metric into a 5-dimensional Lorentz space. This will be of help for
understanding the behaviour of the Regge equations when applied to the
evolution of a 600-cell space section.

The metric in consideration can be written in this way:
$$
\d x^\a\,\d x_\a=-\d t^2+
R^2\left[\d\chi^2+\sin^2\chi\left(\d\vartheta^2+\sin^2\vartheta\,\d\varphi^2
\right)\right],\label{labellag}$$
where $R$ depends only on the universal time $t$.

The variety with metric \ref{labellag} can be embedded into a 5-dimensional
space endowed with the following metric:
$$\d x^\a\,\d x_\a=-\d v^2+\d w^2+\d x^2+\d y^2+\d z^2$$
by putting
$$\allinea{%
 w&=R\cos\chi\cr
 x&=R\sin\chi\sin\vartheta\cos\varphi\cr
 y&=R\sin\chi\sin\vartheta\sin\varphi\cr
 z&=R\sin\chi\cos\vartheta.\cr}\label{coordinate}$$
Equations \ref{coordinate} imply
$$
\d x^\a\,\d x_\a=-\d v^2+\d R^2+
R^2\left[\d\chi^2+\sin^2\chi\left(\d\vartheta^2+\sin^2\vartheta\,\d\varphi^2
\right)\right].\label{gimmersa}$$
In order that metrics \ref{labellag} and \ref{gimmersa} may coincide it is
necessary that
$$\d t^2=\d v^2-\d R^2.\label{dv}$$
If we impose that $R$ is a function only of the time variable $v$ then we have
$$\d R=\frac{\d R}{\d v}\,\d v,$$
so
$$\d t^2=\d v^2-\left(\frac{\d R}{\d v}\right)^{\!2}\d v^2=\funzR\d v^2,$$
from which it follows that $\dot R^2<1$. In this case
$$\d t=\funzR^{\!1/2}\,\d v.\label{dvindt}$$

\noindent So the metric becomes:
$$
\d x^\a\,\d x_\a=-\funzR\d v^2+
R^2\left[\d\chi^2+\sin^2\chi\left(\d\vartheta^2+\sin^2\vartheta\,\d\varphi^2
\right)\right].\label{newg}$$

In the following we will refer to $v$ as ``outer time.'' We will be mainly
interested in the case where $R(v)$ is an even function, so that $v=0$ is an
instant of time symmetry.

\section{The 600-cell and the construction of the initial sandwich
\label{600topo}}*
\subsection{The spatial section\label{sezione}}
We will approximate the hypersphere with the last tridimensional regular
polytope,
the 600-cell. It is a simplicial complex constituted by 120
vertices and 600 tetrahedra. The Friedmann metric has an instant of time
symmetry where we can put a spatial section with a null extrinsic curvature.
In \CdR, for a spatial section whose extrinsic curvature vanishes, the
following relation must hold:
$$\sum_j\eps _{ij}\,l_{ij}=16\pi\,\frac M{120},$$
where the sum is done over all vertices $j$ joined to vertex $i$ by an
edge $[ij]$ of length $l_{ij}$ and defect $\eps_{ij}$. There are 5 regular
tetrahedra around a single edge, and the defect of each edge is:
$$\eps=2\pi-5\,\arccos\!\left(\case13\right)\!.$$
Since every vertex is joined to 12 vertices which make an icosahedron, and the
600-cell is regular, calling $l_0$ its edge, we have
$$M=\frac{90}\pi\,\eps\,l_0.\label{mlfried}$$

In order to have a criterion to compare the results of \CdR\ with those of
\RG, we choose that the 600-cell should approximate a 3-sphere having the same
volume.  Thus the radius $R_0$ of the 3-sphere, in the instant of time
symmetry, will obey the relation
$$2\pi^2\,R_0^3=600\,\frac{l_0^3}{6\sqrt2}$$
or
$$l_0=\zeta\,R_0,$$
where
$$\zeta=\left(\frac{\pi ^2\sqrt2}{50}\right)^{\hsmash{\!1/3}}.$$

\subsection{Creation of the initial sandwich}
We call ``initial sandwich'' the \ST\ between the initial spacelike
section ($\Sigma_0$) and the next one ($\Sigma_1$).

\subsubsection{The classes of vertices:
$\a,\be,\g,\de,\eps$\label{sezvertici}}
\Tuckey\ and \Barrett\ say the 120 vertices of a 600-cell can be subdivided
into 4 classes. In each class there should be 30 vertices not joined to one
another by an edge. So all vertices of a class could be evolved
simultaneously.

But this cannot be true: it can be shown as follows. Let us consider a vertex
$V$: we know it is surrounded by 12 vertices $W_i$ making an icosahedron.
These vertices {\sl cannot\/} be of the same class of $V$ because every vertex
$W_i$ is joined to $V$ by an edge. But in an icosahedron it is not possible to
choose more than 3 vertices not joined to one another, so that these 12
vertices must belong to at least 4 distinct classes.

To overcome this obstacle let us reason as follows. Let us consider~$V$ and
let it be a vertex of class $\a$. Let us try to subdivide the vertices $W_i$
into 4 groups of 3 vertices not joined to one another. We will name these
groups with the letters $\be,\g,\de,\eps$. This can be done by looking at
figure~\ref{icosaedro}.


Let us try to subdivide the vertices of the 600-cell into 5 classes of 24
vertices each. Every class must consist of vertices not joined together. Using
the classification of the vertices of the 600-cell given by \Coxeter, we
find that
$$\eqalign{%
\alpha&=\{A_0,A_{10},A_{20},A_{30},A_{40},A_{50},
B_{7},B_{17},B_{27},B_{37},B_{47},B_{57},\cr
&\qquad
C_{1},C_{11},C_{21},C_{31},C_{41},C_{51},
D_{8},D_{18},D_{28},D_{38},D_{48},D_{58}\}.\cr}$$

The other classes are obtained by adding 2, 4, 6, 8 to the subscripts. It is
also possible to show that the 24 vertices of each class make a 24-cell
regular polytope inscribed in the 600-cell. Furthermore we can see that every
vertex is actually surrounded by 12 vertices equally subdivided among the
other classes. So an $\a$-vertex is joined to 3 $\be$-vertices, 3
$\g$-vertices, 3 $\de$-vertices and 3 $\eps$-vertices. It should be noted that
the 3 vertices of the same class joined to a same vertex are not
equivalent. For example the 3 $\be$-vertices around the vertex $\a$ in
figure~\ref{icosaedro} have different neighbouring vertices. In fact they are
surrounded by three sets (of 5 vertices each), which are not equal to one
another. By the edge $[\a\be_0]$ we will call the edge $[\a\be]$ surrounded by
the set of the following vertices $\{\g,\g,\de,\de,\eps\}$. The edge
$[\a\be_1]$ is surrounded by the set $\{\g,\g,\de,\eps,\eps\}$ and $[\a\be_2]$
by $\{\g,\de,\de,\eps,\eps\}$.  As we shall see later, this inequality among
the three edges $[\a\be_i]$ ($i=0,\dots,2$) will give birth to three different
Regge equations when an $\a$-vertex is going to be evolved.

Because of all these facts the evolution of a 600-cell consists of {\sl
five\/} steps: the evolution of each class of vertices ($\a$, $\be$, $\g$,
$\de$, $\eps$).

\subsubsection{The initial sandwich and the instant of time symmetry
\label{timesym}}
Equation \ref{mlfried} has been found by choosing a spatial section having a
null extrinsic curvature. In the continuum this means that the orthogonal
timelike geodesics, which go out from this section, are locally parallel. In a
lattice too, this characteristic should remain. So we can say that the spatial
section, found by putting $l_0$ equal to the value given by equation
\ref{mlfried}, cannot be a section of time symmetry. In fact if this were
true, the edge lengths in the next section would be smaller: $l'=l_1<l_0$.
Then the timelike edges which are the geodesics of the masses could not be
parallel.

To understand better what happens in this situation let us embed the 600-cell
into the 5-dimensional lorentzian space described in section~\ref{secEmb}.
Since the 600-cell can be inscribed in a 3-sphere, we can look at its
evolution as a sequence of 3-spheres each having a radius $R(v)$, where $v$ is
the outer time of the 5-space. Then we put
$$\a=(v,R\,\vu_\a)$$
where $\vu_\a$ is a unit spacelike 4-vector,
$$\vu_\a=(\cos\chi_\a,\,\sin\chi_\a\sin\vartheta_\a\cos\varphi_\a,\,
\sin\chi_\a\sin\vartheta_\a\sin\varphi_\a,\,
\sin\chi_\a\cos\vartheta_\a),$$
pointing to the direction of $\a$. There will be one such vector $\vu$ for
each vertex of all 600-cell spatial sections. Let us call $\Sigma_0$ the
section corresponding to the instant of time symmetry $v=0$, and $\Sigma_1$
the next one, at time $v_1$, then
$$\allinea{%
&\a  &&=(0,&&R_0\,&&\vu_\a&&)\qquad\be&&=(0,&&R_0\,&&\vu_\be&&)\cr
&\a' &&=(v_1,&&R_1\,&&\vu_\a&&)\qquad\be'&&=(v_1,&&R_1\,&&\vu_\be&&).}
$$
So we have that $[\a\a']$ is not orthogonal to $[\a\be]$, unlike what happens
in the continuum. In fact the straight lines
$$\allinea{%
{\bf x}&=\a+&&p\,&&[\a\a']\cr
{\bf x}&=\be+&&q\,&&[\be\be']}$$
are never parallel and cross for $p=q=R_0/(R_0-R_1)$.

Nevertheless we can make an attempt in order that our lattice may have a
behaviour similar to that of the continuum. Let us now put $\Sigma_0$ and
$\Sigma_1$ symmetric about the instant of time symmetry $v=0$, then
$v_0=-v_1$. Now we have
$$\allinea{%
&\a &&=(-v_1,&&R_1\,&&\vu_\a&&)    \qquad \be &&=(-v_1,&&R_1\,&&\vu_\be&&)\cr
&\a'&&=(\phm v_1,&&R_1\,&&\vu_\a&&)\qquad \be'&&=
(\phm v_1,&&R_1\,&&\vu_\be&&).} $$
Thus we find that
$$\allinea{%
&[\a'\be']&&=[\a\be],\cr
&[\a\a']&&=[\be\be']=(2\,v_1,0,0,0,0),}$$
and $[\a\a']$ is orthogonal to $[\a\be]$. Therefore we obtain
$$[\a'\be]\cdot[\a'\be]=[\a\a']\cdot[\a\a']+[\a\be]\cdot[\a\be]=
  -4\,v_1^2+l_0^2\approx-\tau_0^2+l_0^2,$$
where $\tau_0$ is the interval of proper time between $\Sigma_0$ and
$\Sigma_1$. In fact we have also $\tau_0\approx\Delta v$ because $\dot R=0$
for $v=0$.

Now all timelike edges (geodesics of the mass points) are orthogonal to the
spatial section and parallel to one another, as we would expect if the
extrinsic curvature vanished. Equation \ref{mlfried} does not give the edge of
the time symmetric spatial section, but the edge of two consecutive 600-cells
equal and symmetric about $v=0$. All other spatial sections will also be
pairwise symmetric about $v=0$.

So we know the squared lengths%
\footnote{$^{\smallfonts\dag}$}{We call ``squared length'' of an edge $[ij]$
the real quantity $s_{ij}=[ij]\cdot[ij]$, whereas its length is the
non-negative number $l_{ij}=|s_{ij}|^{1/2}$}
of all edges of the initial sandwich and are able to build it at once without
using any Regge equation.

\subsubsection{The initial sandwich using Regge equations}
To test our assumptions we evolved the section $\Sigma_0$ making use
of the Regge equations for the edges lying between $\Sigma_0$ and $\Sigma_1$,
according to the Sorkin evolutive scheme. In order to allow an easier
comparison between our results and those of \Barrett\ we have chosen a lattice
having
$$\allinea{%
M&=10.202\ldots\;\hbox{ (in arbitrary units)},\cr
l_0&=2.774\ldots,\cr
\tau _0&=0.0102.}$$

In fact the authors considered the Friedmann universe having maximum radius
$R_0$ and mass $M_{*}$ defined by
$$M_{*}\ugual\int\!\varrho(t)\,dV=2\pi^2\,\varrho R^3=\frac{3\pi}4\,R_0.$$
Choosing $M_{*}=10$ they found
$$l_0=\zeta \,R_0=\frac{4\zeta}{3\pi }\,M_{*}=2.774\ldots$$
In the end they found, using equation \ref{mlfried}, the value $10.202\ldots$
for $M$.

We have already remarked that the evolution scheme does not preserve the exact
symmetry of the initial 600-cell. We expect however that, if the evolution is
properly done, the symmetry is approximately respected. In order to test this,
we examined those variables which are bound to remain equal if the symmetry
holds and computed their averages and standard deviations. These variables
are: the squared length $d^2$ of the oblique edge joining $\Sigma_0$ and
$\Sigma_1$ and the squared length ${l'}^2$ of the edge of the new 600-cell
forming the $\Sigma_1$ section.

\noindent We obtained
$$\allinea{%
      &l_0^2   &&=7.693\,799\,901\,383\,04 &&\cr
l_0^2-&\tau_0^2&&=7.693\,695\,861\,383\,04 &&\cr
      &d^2     &&=7.693\,695\,861\,383\,01&\qquad\sigmad&=\10{1.348}{-15}\cr
      &{l'}^2  &&=7.693\,799\,901\,382\,97&\qquad\sigmal&=\10{5.767}{-15}.}$$

So, as we expected, the same value $l_0$ occurs on the first two consecutive
surfaces. Furthermore we see that the relation $d^2=l_0^2-\tau_0^2$ holds
up to the 14th significant figure.

As we already remarked, the other surfaces, to be obtained later, are
symmetric about these two. This last result has been found by \Barrett\ too.
The authors also found a second time-symmetric solution, in which the maximum
spatial length is reached on a single hypersurface and the other hypersurfaces
are symmetric in pairs around it. However this new maximum is slightly larger
than that given by equation \ref{mlfried}. These are all acceptable models and
are indistinguishable, on the scale of time we used, from ours.

\section{The evolution of the initial sandwich by Sorkin
evolutive scheme\label{evoluzioni}}
Given the squared lengths of all the edges lying between and on two spatial
sections $\Sigma_{k-1}$ and $\Sigma_k$, in order to create a new simplicial
sandwich, i.e.\ the \ST\ between $\Sigma_k$ and $\Sigma_{k+1}$, we will
follow five steps.

\iniziaitem
\item{1.}To evolve the vertices of class $\a'$ of $\Sigma_k$. We put a
timelike edge (of negative squared length $s_{\a''\a'}=-\tau_k^2$) joining
$\a'$ to its evolute $\a''$ of $\Sigma_{k+1}$. We link $\a''$ to the 12
neighbouring vertices of $\a'$'s entourage in $\Sigma_k$ through spacelike
edges (to ensure this, $|s_{\a''\a'}|$ should be small enough with respect to
the other edges).\nuovariga
There is one Regge equation for each ``internal'' edge. Then we find Regge
equations around $[\a'_\phz\be_i']$, $[\a'_\phz\g_i']$, $[\a'_\phz\de_i']$,
$[\a'_\phz\eps_i']$ ($i=0\ldots2$) and around $[\a''\a']$. We have as many
equations as unknowns. The problem looks completely determined. Things are not
so easy, however.\nuovariga
As widely shown in literature, on one side we must save the ``lapse and
shift'' freedom; on the other side we expect Regge equations to be not
independent because of the contracted Bianchi identities. Therefore we must
ignore 4 equations and fix 4 of the unknowns by proper lapse and shift
conditions. As condition of lapse we will choose at our convenience
$$s_{\a''\a'}=-\tau_k^2,$$
and as conditions of shift the three constraints
$$\allinea{%
&s_{\a''_\phz\be'_0}&&= s_{\a''_\phz\be'_1} \cr
&s_{\a''_\phz\be'_0}&&= s_{\a''_\phz\be'_2} \cr
&s_{\a''_\phz\g'_0} &&= s_{\a''_\phz\g'_1}.}$$
We then cancel the 4 equations for $[\a''\a']$, $[\a'_\phz\be'_1]$,
$[\a'_\phz\be'_2]$, $[\a'_\phz\g'_1]$.
\nuovariga
As already remarked in subsubsection~\ref{sezvertici}, equations over edges of
the same class, e.g.\ $[\a'_\phz\de'_i]$ are different. In fact it could be
seen that $[\a'_\phz\de'_0]$ is linked to 8 vertices (one $\de$, one $\eps$,
two $\be'$, two $\g'$, one $\eps'$ and one $\a''$) whereas $[\a'_\phz\de'_1]$
is linked to 9 vertices (one $\de$, two $\eps$, two $\be'$, one $\g'$, two
$\eps'$ and one $\a''$).
\nuovariga
There are still the following 9 unknowns:
$$s_{\a''_\phz\be'_0},\; s_{\a''_\phz\g_0'},\; s_{\a''_\phz\g_2'},\;
s_{\a''_\phz\de_i'},\;s_{\a''_\phz\eps_i'}.$$
Nevertheless, we can say that the 12 slanting edges belong to 4 different
classes, corresponding to the class of the second vertex. It is natural to
require, in order to save the symmetry, that edges belonging to the same class
be of equal length. So the unknowns reduce to 4:
$$s_{\a''\be'},\;s_{\a''\g'},\;s_{\a''\de'},\;s_{\a''\eps'},$$
after putting equal all connections $[\a''\de']$, all connections
$[\a''\eps']$ and $s_{\a''_\phz\g'_0}=s_{\a''_\phz\g'_2}$. So the equations
must reduce to four, for instance:
$$[\a'_\phz\be'_0],\;[\a'_\phz\de'_0],\;[\a'_\phz\g'_0],\;[\a'_\phz\eps'_0].$$
In the following we shall verify that the other equations we have not used are
satisfied anyhow.

\item{2.}To evolve the vertices of class $\be'$.
We join $\be'$ to $\be''$ and link $\be''$ with the 3 vertices of class $\a''$
of its entourage in $\Sigma_{k+1}$ and with the 9 vertices (of classes $\g'$,
$\de'$ and $\eps'$) of $\be'$'s entourage in $\Sigma_k$. By symmetry, we may
reduce the new edge lengths to the following:
$$s_{\be''\g'},\;s_{\be''\de'},\;s_{\be''\eps'},\;s_{\a''\be''},$$
since we put as lapse $s_{\be''\be'}=-\tau_k^2$. The equations are
$$[\be'\g'],\;[\be'\de'],\;[\be'\eps'],\;[\a''\be'],$$
neglecting the equation for $[\be''\be']$ because of Bianchi identities.

\item{3.}To evolve the vertices of class $\g'$.
We join $\g'$ to $\g''$ and link $\g''$ with the 6 vertices (of classes $\a''$
and $\be''$) of its entourage in $\Sigma_{k+1}$ and with the 6 vertices (of
classes $\de'$ and $\eps'$) of $\g'$'s entourage in $\Sigma_k$. Now we have
five kinds of internal edges: $[\g''\g']$, $[\a''\g']$, $[\be''\g']$,
$[\g'\de']$ and $[\g'\eps']$. These give rise to five distinct Regge
equations. But because of the Bianchi identity only four of these can be used
to find four lengths (neglecting the one for the timelike edge), whereas
$s_{\g''\g'}$ must be freely given, like $s_{\a''\a'}$ and put equal to
$-\tau_k^2$. Therefore we take as unknowns
$$s_{\g''\de'},\;s_{\g''\eps'},\;s_{\a''\g''},\;s_{\be''\g''}.$$

\item{4.}To evolve the vertices of class $\de'$.
We join $\de'$ to $\de''$ and link $\de''$ with the 9 vertices (of classes
$\a''$, $\be''$ and $\g''$) of its entourage in $\Sigma_{k+1}$ and with the 3
vertices of class $\eps'$ of $\de'$'s entourage in $\Sigma_k$. By symmetry, we
impose that the new edge lengths are:
$$s_{\de''\eps'},\;s_{\a''\de''},\;s_{\be''\de''},\;s_{\g''\de''},\;
s_{\de''\de'}.$$
We have only five distinct Regge equations, around $[\de''\de']$,
$[\a''\de']$, $[\be''\de']$, $[\g''\de']$ and $[\de'\eps']$. One of these,
that for the timelike edge, will not be utilized because of the Bianchi
identity. So the remaining four can be used to find four lengths, whereas
$s_{\de''\de'}$ is given, like $s_{\a''\a'}$ and equal to $-\tau_k^2$.

\item{5.}To evolve the vertices of class $\eps'$.
We join $\eps'$ to $\eps''$ and link $\eps''$ with the 12 vertices (of classes
$\a''$, $\be''$, $\g''$ and $\de''$) of its entourage in $\Sigma_{k+1}$. Now
we have five kinds of internal edges: $[\eps''\eps']$, $[\a''\eps']$,
$[\be''\eps']$, $[\g''\eps']$ and $[\de''\eps']$. These give rise to five
distinct Regge equations. But because of the Bianchi identity only four of
these can be used to find four lengths (neglecting that for the timelike
edge), whereas $s_{\eps''\eps'}$ must be freely given, like $s_{\a''\a'}$ and
put equal to $-\tau_k^2$. Therefore we take as unknowns
$$s_{\a''\eps''},\;s_{\be''\eps''},\;s_{\de''\eps''},\;s_{\g''\eps''}.$$
\par\fineitem

It should be noted that every evolution consists in solving five systems of 4
equations in 4 unknowns each. Furthermore the simplicial sandwich does not
have the same grade of symmetry as the initial spatial section, where all
vertices were equivalent. The evolutive algorithm has broken the symmetry
among vertices $\a$, $\be$, $\g$, $\de$, $\eps$.\nuovariga
The evolution of $\Sigma_k$ is now finished: the new sandwich has been
built up.

\subsection{Results of the evolution}
We evolved the initial 600-cell by imposing that the edges of the same
type, like $[\a'_\phz\be'_i]$, were be all equal. In this way only four
unknowns and four equations remained when each vertex was evolved. Actually
the different variables should be only two, since symmetry requires that all
edges $[v''w']$ and $[v''w'']$ are equal as well. This was done by \Barrett.
By allowing the edges joining vertices belonging to two different classes to
be possibly different one from another, we introduce new degrees of freedom in
the lattice. We have verified that this enhancement of freedom causes no
problem in the evolution.

\subsubsection{Behaviour of $R_k$ as a function of the universal time $t$}
Using the evolutive algorithm we evaluated, for each section, the edge (or
better the mean edge) of the 600-cell, $l_k$. Then we calculated the quantity
$$R_k=\frac{l_k}\zeta,$$
where $R_k$ represents the radius of the 3-sphere equivalent to the 600-cell
of the $k$-th spatial section ($\Sigma_k$) generated by the $k$-th evolution.
We have that $l_k=l_k(t)$, where
$$t=\sum_{i=0}^{k-1}\tau_i,$$
$\tau_i$ is the interval of proper time between $\Sigma_i$ and $\Sigma_{i+1}$,
and $\tau_0=0.0102$. As better explained in the appendix, we did not choose
equal proper time intervals. Instead we took $\tau_i=\rho\,l_i$, with
$\rho\approx\10{3.68}{-3}$.

The results of this evolution can be seen in figure~\ref{rfunzionet}.


\noindent Figure \ref{rfunzionet} shows that the evolution stops at a finite
value of the radius ($R_{\rm m}\approx1.2$) which is approximately $\frac14$
of the initial value ($R_0=4.244\ldots$). For $R<R_{\rm m}$ Regge equations
give no solution.

As it can be seen in the appendix our evolution steps were done with a
constant interval of conformal time: equation \ref{deta} says that
$\Delta\eta\approx\10{2.4}{-3}$. Actually the critical point occurs after
$k_{\rm m}\approx 750$ evolutions, or $\eta_{\rm m}\approx 1.8$. But, since
$\eta$ should reach $\pi$, we should have done a number
$$k_{\rm M}^{}=\frac\pi{\Delta\eta}\approx1300$$
of evolutions. This behaviour of our ``platonic'' polytope was also found by
\Barrett\ and \Brewin\ who also studied not-simplicial complexes that,
compared to the 600-cell, better approximated the 3-sphere. For these
polyhedra the graph of $R(t)$ is nearer to the cycloidal one of the Friedmann
metric and the point of stop is placed in $\eta>\eta_{\rm m}$.

\subsubsection{Behaviour of the two $\sigma$}
During the evolution, the two $\sigma$ ($\sigmad$, $\sigmal$) remain below
$10^{-13}$. In proximity of the critical point these variables grow up
quickly. This means that the lattice is twisting more and more.

\subsubsection{Bianchi identity}
We calculated, for each evolution, the residual $f_{\a''\a'}$ of the Regge
equation for the timelike edge $[\a''\a']$,
$$f_{\a''\a'}=\sum_k\eps _{\a''\a'k}\,\frac{\partial A_{\a''\a'k}}{%
\partial l_{\a''\a'}}-\frac\pi{15}\,M.$$
This equation had not been utilised to build up the spatial section
$\Sigma_{k+1}$ because of the simplicial version of the contracted Bianchi
identity, i.e.~the energy conservation in the form of the Kirchhoff-like law.
The smaller $f_{\a''\a'}$, the better this identity is respected. It should be
noted that the Kirchhoff-like law always exactly holds because the current of
matter along the timelike edges remains the same after each evolution.
Instead, Regge equations represent the Einstein tensor only approximately
\tonde{\Barret}. Our results can be seen in figure~\ref{bianchi}: we see that
$f_{\a''\a'}$ is going to ``explode'' in proximity of the critical point.

The other equations, which are not utilized during the evolution, are all over
slanting and orizontal edges. We found that they are satisfied with greater
accuracy, since their residuals are all less than $\10{2}{-15}$.


\section{Analysis of the critical point\label{analisi}}
What is the nature of the critical point? \Brewin\ gives a partial answer
by noting that the following relation occurs:
$$\frac{\Delta l}\tau\approx1,$$
i.e.~the collapse becomes so fast that the vertical edges are to become
spacelike. Unfortunately in our model the above fraction reaches a not easily
interpretable value (approximately 2.6). So what is the true meaning of this
condition, and how can we continue the iterations (if possible)?

To answer these questions, it is useful to study more deeply the
characteristics of the homogeneous and isotropic metric with topology (and
geometry) $S^3$. The 5-dimensional embedding described in section \ref{secEmb}
will be of use here.

\subsection{Singularities}
In order to study the singularities of metric \ref{newg} we can compute
the quadratic Riemann invariant:
$$Q = R^{\a\be\g\de}\,R_{\a\be\g\de}=
\frac{12}{R^2\bigl(1-\dot R^2\bigr)^2\plico}
\left[\frac{\ddot R^2}{\bigl(1-\dot R^2\bigr)^2\plico}+\frac1{R^2}\right].
\label{riem}$$
In fact a necessary and sufficient condition for a singularity is that $Q$
becomes infinite. A glance to equation \ref{riem} shows that the singular
points are those values of $v$ where $\valass{\dot R}=1$ or $R=0$ or
$\valass{\ddot R}=+\infty$. Let us discuss the possible cases.

\iniziaitem
\item{1.}{\sl The points\/} $R=0$. These coincide with the infinite
contraction of the universe when the volume of the spatial section goes to 0.
In these points, as will be shown later, density reaches infinite values.
By a volume-vanishing (VV) singularity we will mean one of these points.

\item{2.}{\sl The points\/} $\dot R=\pm1$. If we are not in the first case we
can say that the radius of the space section increases or decreases with speed
tending to that of light. If one tries to extend the manifold beyond these
points, and if $\valass{\dot R}$ increases further, the metric becomes
positively defined. So it can no longer represent a manifold of \RG. This is
tantamount to saying that the spacelike sections are not causally connected:
the points of two different spatial sections can be joined only by spacelike
geodesics. So these points will be said to be causality-breaking (CB)
singularities. Actually, in a rigorous sense a CB singularity is a border of
the manifold, and the extension is meaningless.

\item{3.}{\sl The points\/} $\ddot R=\pm\infty$. If we are not in the previous
two cases the singular behaviour is shown by the geodesics deviation, which
becomes infinite. We will refer to these points as geodesics-deviation (GD)
singularities.
\par\fineitem

\noindent These singularities can also occur together in the same point: an
example is the hypothetical universe whose radius has the form
$R(v)=v-A\,v^{3/2}$ (the critical point $v=0$ is a VV-CB-GD singularity).

\subsection{Study of the stress-energy tensor}
From Einstein equations for the closed Robertson--Walker metric we have for
the diagonal components of $T_{\a\be}$ in an orthonormal basis:
$$\eqalign{
\frac{8\pi}3\,\r&=\frac1{R^2}\left(\frac{\d
R}{\d t}\right)^{\!2}+\frac1{R^2}\cr -4\pi\,p&=\frac1R\,\frac{\d^2R}{\d t^2}+
          \frac1{2R^2}\left(\frac{\d R}{\d t}\right)^{\!2}+\frac1{2R^2}.}$$
By expressing the time derivatives in terms of the outer time we find
$$\eqalign{%
\r&=\frac3{8\pi}\frac1{R^2\bigl(1-\dot R^2\bigr)\plico}\cr
p&=-\frac1{8\pi}\frac{2R\ddot R+1-\dot R^2}
                     {R^2\bigl(1-\dot R^2\bigr)^2\plico}.}\label{rhop}$$

\subsubsection{Range of validity of the stress-energy tensor}
It is well known that the stress-energy tensor must satisfy
$$\r \geq |p|.$$
In our case
$$-2\le\frac{R\ddot R}{1-\dot R^2\plico}\le1.
\label{rhogeqp}$$
In general it can happen that condition \ref{rhogeqp} be not satisfied for
values of $v$ for which the metric still has a ``geometrical sense.'' In fact
a CB or GD or CB-GD singularity does not satisfy equation \ref{rhogeqp}.
Instead a VV point presents no problem. It should be noted that we cannot
become aware of these new ``physical'' matter-limiting (ML) points, wherein
$\r=|p|$, by looking only at the expression of the metric.

\subsection{Friedmann metric}
Since $\d t=R\,\d\eta$, from equation \ref{dv} we obtain
$$
v=\int\!\left[R^2-(\d R/\d\eta)^2\right]^{\!1/2}\d\eta.$$
For the Friedmann metric we have:
$$v=2R_0\,\sin\frac\eta2$$
and
$$R=R_0\cos^2\frac\eta2=R_0-\frac{v^2}{4R_0}.$$
Metric \ref{newg} takes the following form:
$$
\d x^\a\,\d x_\a=-\frac R{R_{\hsmash0}}\>\d v^2+
R^2\left[\d\chi^2+\sin^2\chi\left(\d\vartheta^2+\sin^2\vartheta\,\d\varphi^2
\right)\right].$$

Calling $v_c$ the value of $v$ where $\dot R=-1$ and $\vom$ the positive value
of $v$ where $R=0$, for the Friedmann metric these two singularities occur
{\sl at the same time\/}, i.e.~$\vom=v_c=2R_0$. So, in the 5-space, in
correspondence of the big crunch the speed of contraction equals the speed of
light, i.e.\ the big crunch is a VV-CB point. This behaviour is shared by all
matter satisfying $p=w\,\r$, where $w$ is a constant, $-\frac13<w\le1$, since
then $\dot R^2 = 1 - A^2\,R^{1+3w}$, $A$ a constant. Seeing that $\ddot R =
-\frac12\, A^2\, (1+3w)\,R^{3w}$, in Zel'dovich's interval, $0\le w\le1$, no
GD singularity is present beside the other two.

\section{The problem of the 600-cell\label{problema}}
Let us now reconsider equation \ref{dv}. For each evolutive step we know $\Delta
t=\tau$ and $\Delta R=\Delta l/\zeta$. Then we can find $\Delta v$ by
calculating $\Delta v \approx \tondemat{\Delta t^2+\Delta R^2}^{\!1/2}$.
So we determined $R$ as a function of $v$ (figure~\ref{rfunzionev}).


\noindent We have also made a picture of $\Delta R/\Delta v\approx\d R/\d v$
as a function of $R$ in figure~\ref{dvfunzionedr}.


We can see that this derivative in the last iterations tends, in absolute
value, to 1. In this sense the 600-cell meets a CB singularity before
its volume vanishes. So we expect that for the 600-cell $v_c<\vom$. This
leads us to believe that there is no point in trying to continue the evolution
beyond this point, perhaps by making all edges spacelike as suggested by
\Barrett.

\section{Study of a ``new'' class of solutions of Einstein equations
\label{nuovo}}
We have just seen that the behaviour of the evolution for the 600-cell looks
somewhat different from the Friedmann solution, in its reaching a
CB singularity before the vanishing of $R$. So it appears expedient to
study a generalisation of Friedmann metric, also having that property.
For the 600-cell an empirical fit of $R(v)$ shows that a good fit is given by
$$R(v)=R_0-\frac{a^2v^2}{4R_0},\qquad{\rm with}\qquad a^2\approx1.128,$$
whereas $a=1$ is the Friedmann metric.

It is useful to re-define
$$\allinea{%
&v_c&&\ugual\frac{2R_0}{a^2},\cr
&\vom&&\ugual\frac{2R_0}a=a\,v_c,\cr}$$
so
$$\eqalign{%
R&=\frac1{2v_c}\left(v_\Omega^2-v^2\right)\cr
\dot R&=-\frac v{v_{\hsmash c}}\cr
\ddot R&=-\frac1{v_{\hsmash c}}\>.\cr}$$

\iniziaitem
\item{\bu}For $0<a<1$ we have that $-\vom<v<\vom$ and the singularity is of
VV-type.

\item{\bu}For $a>1$ we have that $-v_c<v<v_c$ and the singularity is of
CB-type.

\item{\bu}For $a=1$ the solution is the Friedmann metric. So this case
is a watershed between two different behaviours of the general metric.
\par\fineitem

\noindent It is not difficult to find the expression of~$t$ as a function
of~$v$:
$$t=\case12\,v_c\arcsin(v/v_c)+\case12\,v\left[1-(v/v_c)^2\right]^{\!1/2}$$
and to show that the graph of~$R(t)$ is still a cycloid, like for Friedmann
metric, but scaled in~$R$ by a factor~$1/a^2$ and translated upwards by~$\frac12\,
v_c\left(a^2-1\right)$.

\subsection{Behaviour of the metric}
Condition \ref{rhogeqp} implies that the following inequality must
be satisfied:
$$v^2\leq\tilde v^2\ugual v_c^2-\case13\,v_c^2\left|a^2-1\right|.$$
If $a<1$ this condition does not carry any problem because $\tilde v^2 >
v_\Omega^2$. If $a>1$ the ML point $\tilde v$ is the first one the metric
meets in its evolution. In fact
$$\tilde v^2=v_c^2-\case13\,v_c^2\left(a^2-1\right)<v_c^2.$$
Note that in order to have $\tilde v^2\geq0$ the value of $a^2$ must not
exceed 4. So the conditions for the stress-energy tensor reduce the interval
of definition of $a$ to $0<a\leq2$. If $a<1$, the metric ends in a VV
singularity reached the more softly the smaller the value of $a$. If $a>1$,
the metric reaches a ML point before the CB one. It should be also pointed out
that the Friedmann metric ($a=1$) is more ``pathological'' because its VV
singularity is also a CB point, since $\vom=v_c$.

For this metric it is also possible in principle to write down an
equation of state of matter, but of course we do not attach a physical meaning
to this result.

\section{Conclusions\label{conclusioni}}
We have seen that the evolution of the 600-cell does not describe the
Friedmann universe well. Instead we can think of it as the evolution of a
different type of matter. But where does such a difference come from? In our
opinion, one should consider the following two points as playing an important
role in answering this question.

First, the 600-cell is a good approximation to a 3-sphere, but it is still too
``rough.'' For every approximating method, \CdR\ included, the smaller the
step interval the better the fit. When this interval reduces, the two
solutions (numerical and analytical) tend to coincide. In \CdR, in order
to get a better approximation, one should increase the number of tetrahedra of
each space section, i.e.\ the 600-cell should be substituted by another
non-regular polytope nearer to a 3-sphere \tonde{\Brewin}.

The second aspect is deeper since it is closely related to the grounds of the
\CdR\ itself. We have seen that Regge equations give rise to a different
evolution of the universe, ending in a CB point. This happens because they are
only an approximation of Einstein equations. The point has been discussed by
\BrewinG. It can be said that Regge equations are a good approximation to
differential equations rather different from Einstein's. It is wholly possible
that such equations have solutions which behave in a qualitatively different
way from those we are looking for: for instance, the approximate solution may
exhibit singularities not belonging to the correct solution. Nevertheless, it
remains true that the smaller the truncation error, the closer the behaviour
of the numerically approximated equation to that of the exact solution.

Thus an idea comes to mind: could we modify Regge equations in order to have a
smaller truncation error? It has been shown \apritonda see \BrewinG,
\MillerG\chiuditonda\ that the solutions of \CdR\ are, in general, expected
to be second order accurate approximations to the corresponding continuum
solutions, i.e.\ the truncation error for lattices obtained by \CdR\
is $O\tondemat{l^3}$. Thus a research program could be: to find a modification
of Regge equations giving a truncation error $O\tondemat{l^4}$ or better.

\Appendix{Numerical topics}
We did not use a fixed $\tau_k$ for all evolution steps. Instead we imposed a
constant ratio between $\tau_k$ and $l_k$. This in order to ensure that the
Courant criterion continues to be respected also when $l_k$ becomes very
small. We remember that the 600-cell should approximate a Friedmann universe
having the same volume. So the radius $R_k$ of the $k$-th 3-sphere is linked
to the edge $l_k$ of the 600-cell of the spatial section $\Sigma_k$ of the
$k$-th evolutive step by the relation $l_k=\zeta\,R_k$ ($\zeta$ defined in
subsection~\ref{sezione}). Then
$$\frac{\tau_k}{l_k}=\rho\ugual\frac{\tau_0}{l_0}\approx\10{3.68}{-3}.$$
This implies that
$$\Delta\eta\approx\frac{\Delta t}R=\frac{\tau_k}{R_k}=\zeta\,\rho
\approx\10{2.4}{-3}.\label{deta}$$

To solve Regge equations, a system of non linear equations, we use the
Newton--Raphson method. To begin the iterations we have to give values to the
unknowns in order to start the iterations. We call them ``trial values'' and
they are marked by a tilde.

To give a trial value to the variables $\sqlk{}{\a''v'}$ (the squared
lengths of the oblique edges that join $\a''$ to the vertices $v'\neq\a'$ of
$\Sigma_k$) we can use the following argument. Keeping the notations of
subsubsection~\ref{timesym} on, we can write
$$\eqalign{
\sqlk{\tilde}{\a''v'}&=
[\a'\a'']\cdot[\a'\a'']+[\a'\be']+[\a'\be']-
2\,[\a'\a'']\cdot[\a'\be'] = \cr
&=-\tau_k^2+l_k^2-2\,[\a'\a'']\cdot[\a'\be'],}$$
where
$$\eqalign{
[\a'\a'']\cdot[\a'\be']&=
\tondemat{\Delta v_k,(R_{k+1}-R_k)\,\vu_\a}\cdot{}
\tondemat{0,R_k\,(\vu_\be-\vu_\a)}=\cr
&= R_k\,(R_{k+1}-R_k)(\vu_\a\cdot\vu_\be-1)
\approx-\frac{\d R}{\d t}(t_k)\,\tau_k\,R_k\,\xi,}$$
being $\tau_k$ the interval of universal time between the sections
$\Sigma_k$ and $\Sigma_{k+1}$, $R_k$ the radius of the 3-sphere circumscribed
to the 600-cell and
$$\xi\ugual1-\vu_\a\cdot\vu_\be=\frac{3-\sqrt5}4,$$
found by calculating the angle between two vectors pointing to any two
neighbouring vertices of the 600-cell \tonde{\Coxeter}. Furthermore we can see
that $l_k=\phi\,R_k$, where $\phi$ is the golden ratio,
$\frac12\,(\sqrt5-1)$. So we are able to write
$$\sqlk{\tilde}{\a''v'}= -\tau_k^2 + l_k^2-
2\,\frac\xi\phi\,\tau_k\,l_k\,\bigl(l_0/l_k-1\bigr)^{1/2}.$$

Once the vertices $\a'$ have been evolved we put the quantity
$\sqlk{}{\a''v'}$ as trial value for all other oblique edges,
$\sqlk\tilde{\be''v'}$, $\sqlk\tilde{\g''v'}$, $\sqlk\tilde{\de''\eps'}$. At
the moment of the evolution of the vertex $\be'$ we have to give another trial
value to the unknown $\sqlk{}{\a''\be''}$.

In the Friedmann metric we would have
$$R(\eta)=R_0\cos^2(\eta/2)$$
so, since $l=\zeta\,R$, expanding to the second order in $\tau$, we can
calculate an approximate value for the edge of the 600-cell of the section
$\Sigma_{k+1}$. In fact we have
$$\tilde l_{k+1} = l_k - \tau_k\,\zeta\,\bigl(l_0/l_k-1\bigr)^{1/2} -
\case14\,\zeta^2\,l_0\,\bigl(\tau_k/l_k\bigr)^2.$$
After the evolution of $\be'$-vertices the quantity $\sqlk{}{\a''\be''}$ is
determined and we use it to give the trial values for all variables
representing the edge of the 600-cell of the section $\Sigma_{k+1}$, i.e.\
$\sqlk\tilde{\g''v''}$, $\sqlk\tilde{\de''v''}$, $\sqlk\tilde{\eps''v''}$.

\references
\refjl{Barrett J W 1986}{\CQG}{3}{203--6}

\refjl{Barrett J W, Galassi M, Miller W A, Sorkin R D,
Tuckey P A and Williams R M 1997}{Int. J. Theor. Phys.}{36}{815--39}

\refjl{Brewin L C 1987}{\CQG}{4}{899--928}

\refjl{Brewin L C and Gentle A P}{{\rm gr-qc/0006017}}{}{}

\refjl{Collins P A and Williams R M 1973}{\PR\ {\rm D}}{7}{965--71}

\refbk{Coxeter H S M 1973}{Regular Polytopes}{(New York: Dover
Publications, Inc.) pp~247--50}

\refjl{Gentle A P and Miller W A 1998}{\CQG}{15}{389--405}

\refjl{Sorkin R D 1975}{\PR\ {\rm D}}{12}{385--96}

\refjl{Tuckey P A 1993}{\CQG}{10}{L109--13}

\ListaFigure

\centerline{\epsfbox{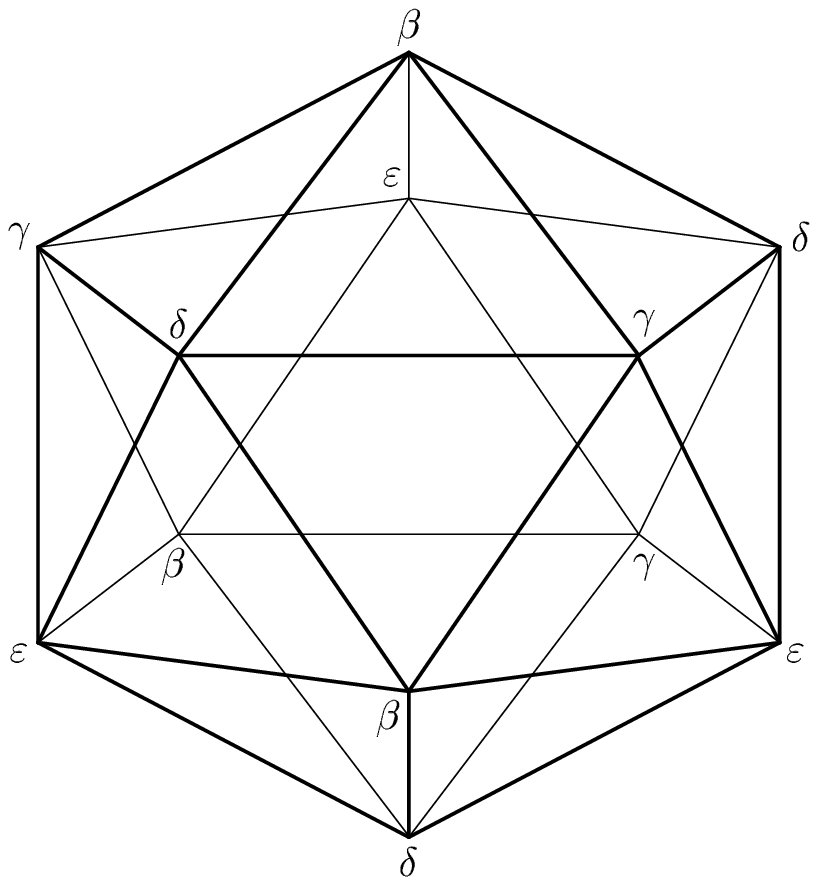}}\bigskip 
\figure{Vertices around an $\a$-vertex\label{icosaedro}}
\vskip2.2truecm

\centerline{\epsfbox{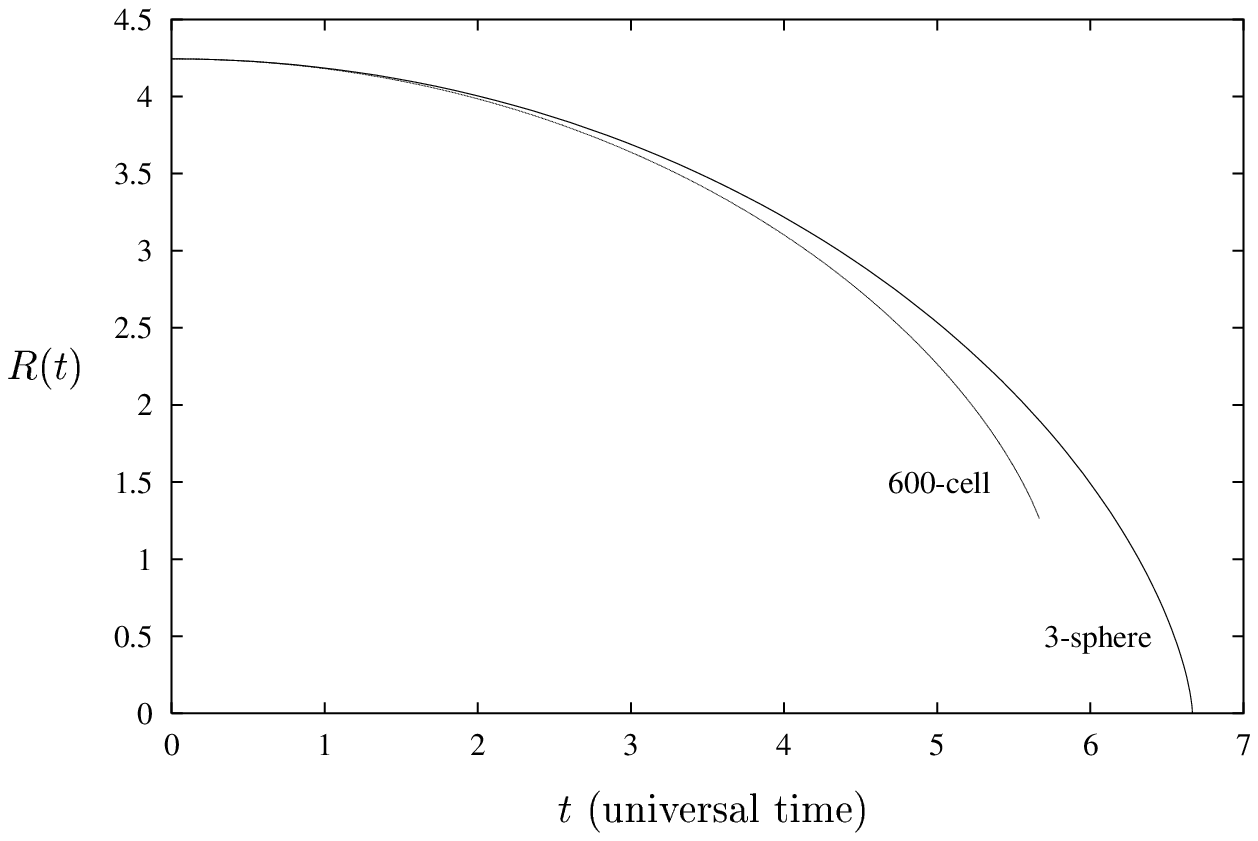}}\bigskip 
\figure{Evolution of the radius $R$ as a function of the universal time $t$
according to \CdR\ and \RG\label{rfunzionet}}

\vfill\eject\null\bigskip

\centerline{\epsfbox{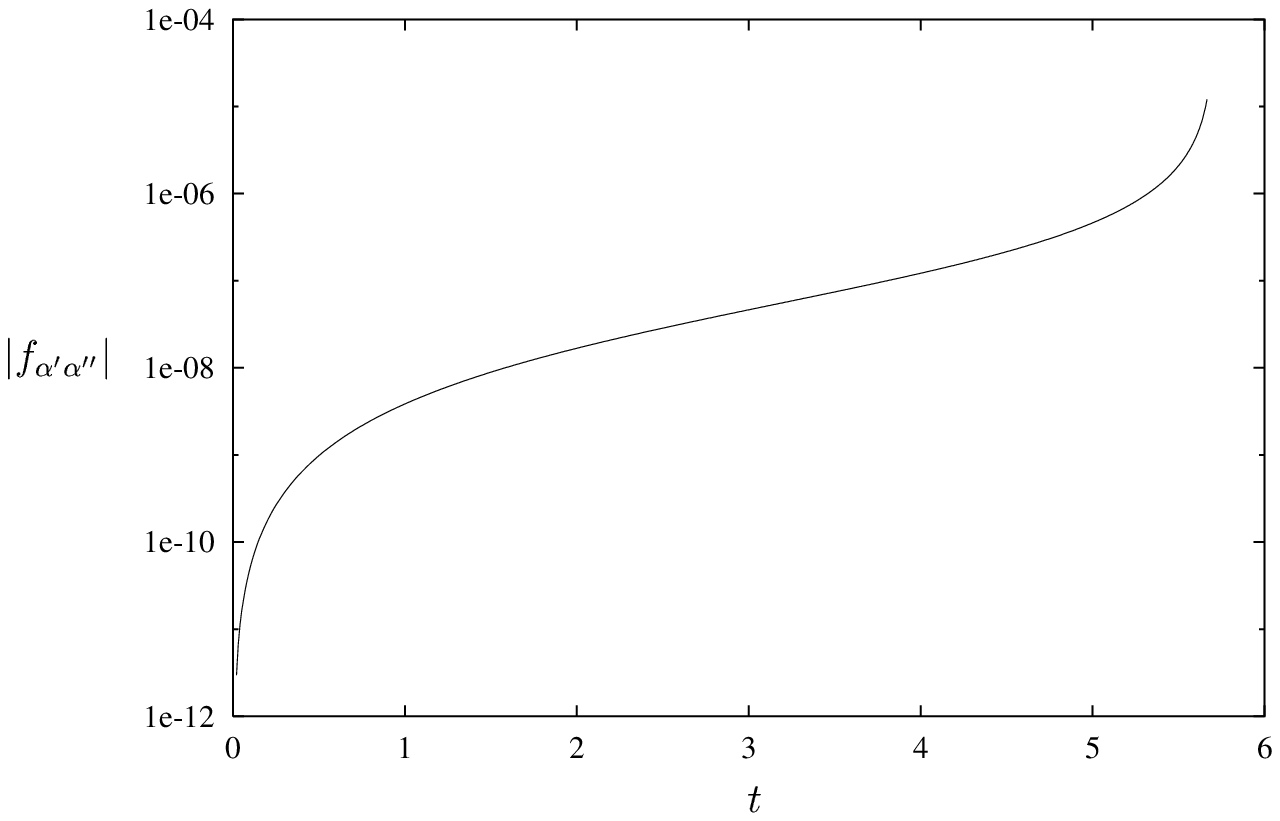}}\bigskip 
\figure{Residual $f_{\a''\a'}$ vs.\ $t$\label{bianchi}}
\vskip1.5truecm

\centerline{\epsfbox{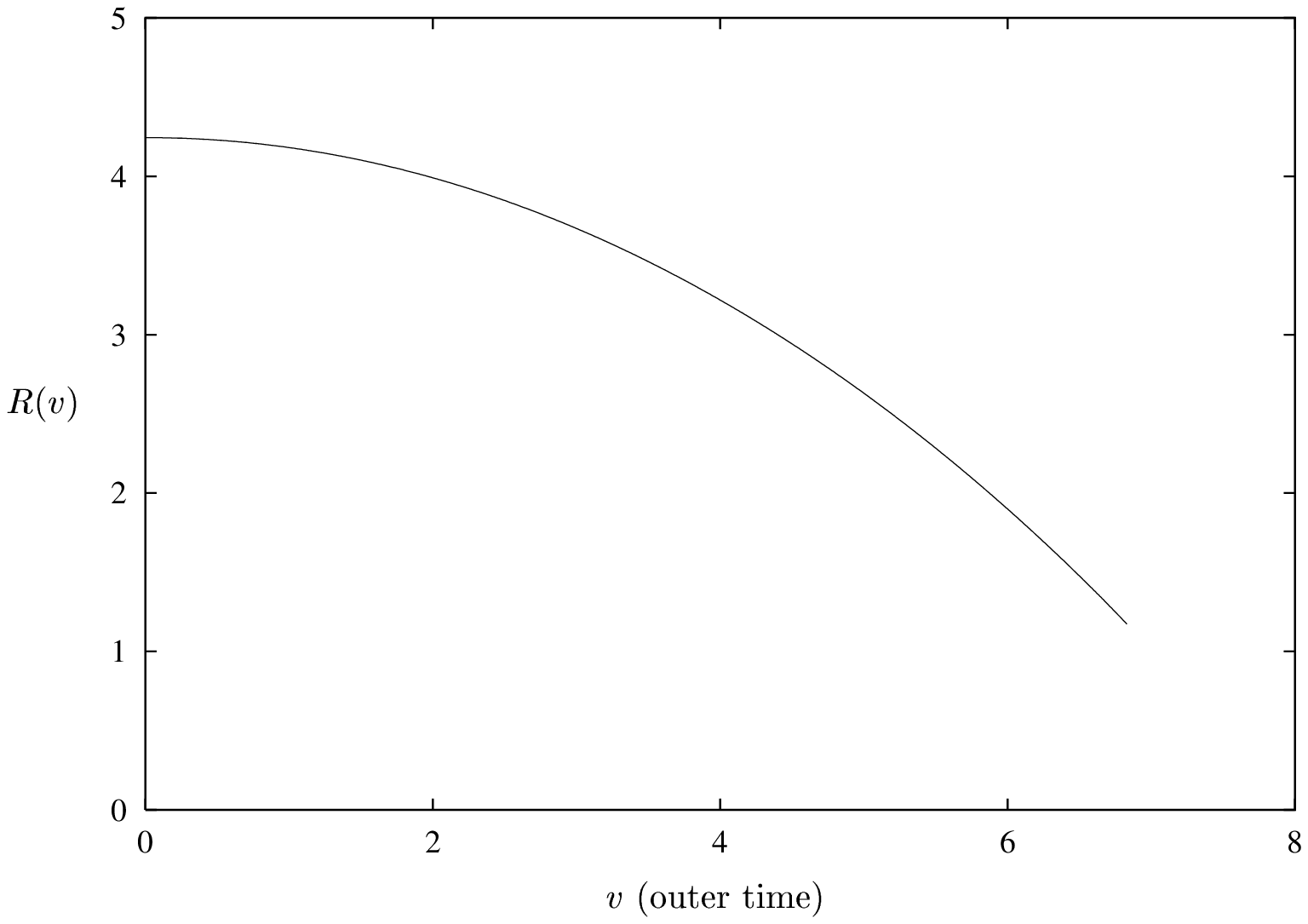}}\bigskip 
\figure{Behaviour of $R$ as a function of $v$.\label{rfunzionev}}

\vfill\eject\null\bigskip

\centerline{\epsfbox{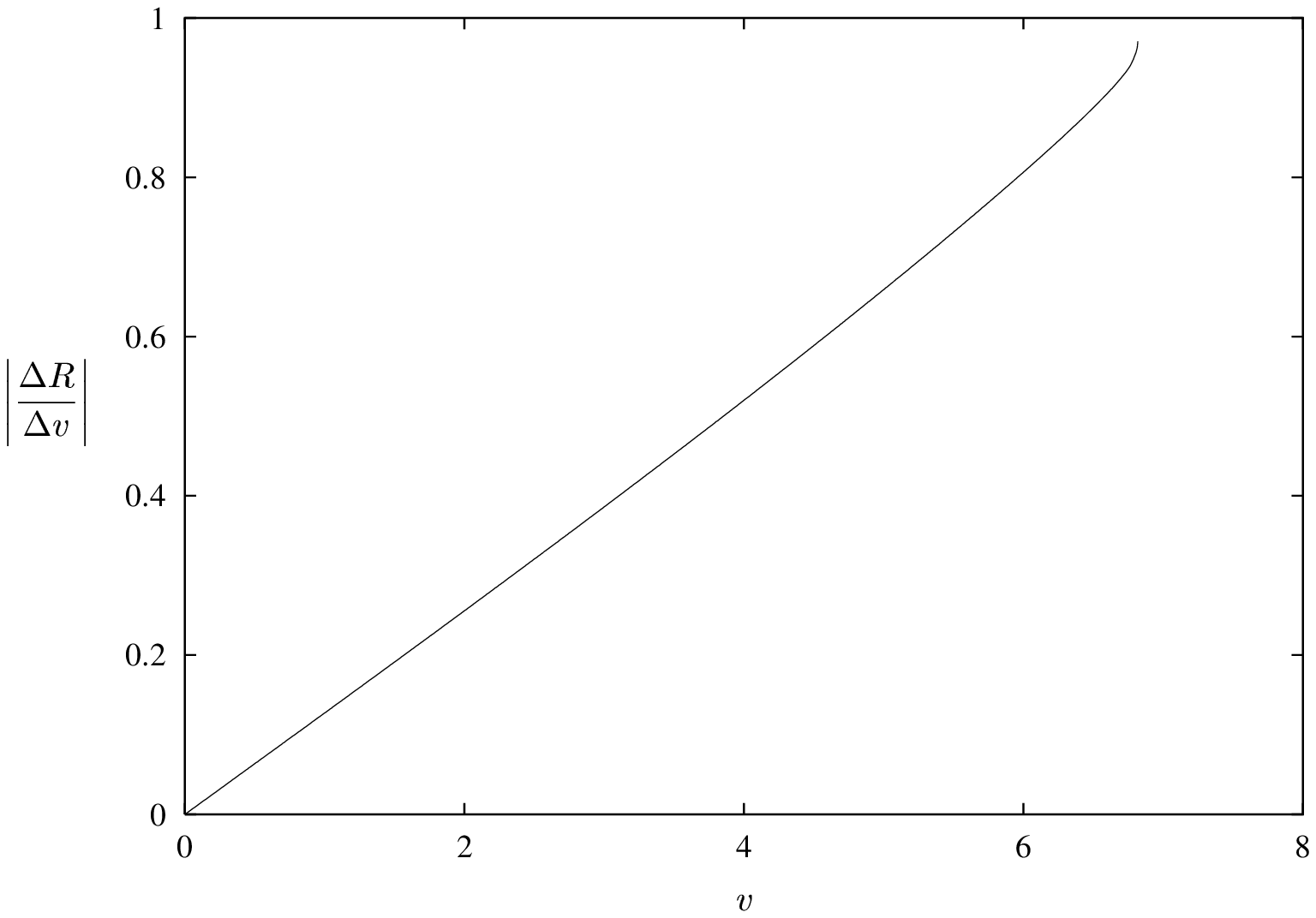}}\bigskip 
\figure{Behaviour of $|\Delta R/\Delta v|\approx \valass{\dot R}$ as a
function of $v$ for the 600-cell\label{dvfunzionedr}}
\bye